\documentclass[letterpaper, aps, tightenlines, nofootinbib, 11pt]{article}
\pdfoutput = 1

\usepackage{euscript}
\usepackage{amssymb}
\usepackage{amsfonts}
\usepackage{amsbsy}
\usepackage{epsfig}
\usepackage{amsthm}
\usepackage{amscd}
\usepackage{amstext}
\usepackage{verbatim}
\usepackage{amsmath}
\usepackage{cancel}
\usepackage{capt-of}
\usepackage{empheq}

\usepackage[pdftex]{hyperref}

\hypersetup{
	colorlinks=true,
	linktoc=page,
	linkcolor=blue,
	citecolor=blue,
	urlcolor=blue
	}

\def\ben{\begin{equation}}
\def\een{\end{equation}}

\let\a=\alpha  \let\g=\gamma \let\d=\delta 
   
\let\l=\lambda     \let\r=\rho

\let\pa=\partial
\def\be{\begin{equation}}
\def\ee{\end{equation}}
\def\beq{\begin{equation}}
\def\eeq{\end{equation}}
\def\ba{\begin{array}}
\def\ea{\end{array}}

\def\dalemb#1#2{{\vbox{\hrule height .#2pt
       \hbox{\vrule width.#2pt height#1pt \kern#1pt
               \vrule width.#2pt}
       \hrule height.#2pt}}}

\newcommand{\bea}{\begin{eqnarray}}
\newcommand{\eea}{\end{eqnarray}}
\newcommand{\tr}{{\rm tr} }
\newcommand{\Tr}{{\rm Tr} }

\def\vep{{\varepsilon}}

\def\Z{{{\Bbb Z}}}

\def\ocal{{\mathcal{O}}}

\newcommand{\bra}[1]{\left\langle #1 \right|}
\newcommand{\ket}[1]{\left|#1\right\rangle}

\newcommand{\nn}{\nonumber \\}

\begin{document}

\begin{center}

{ \Large {\bf
Matrix Quantum Mechanics from Qubits
}}

\vspace{1cm}

Sean A. Hartnoll, Liza Huijse and Edward A. Mazenc

\vspace{1cm}

{\small
{\it Department of Physics, Stanford University, \\
Stanford, CA 94305-4060, USA }}

\vspace{1.6cm}

\end{center}

\begin{abstract}

We introduce a transverse field Ising model with order $N^2$ spins interacting via a nonlocal quartic interaction.
The model has an $O(N,\Z)$, hyperoctahedral, symmetry. We show that the large $N$ partition function admits a saddle point in which the symmetry is enhanced to $O(N)$. We further demonstrate that this `matrix saddle'
correctly computes large $N$ observables at weak and strong coupling. The matrix saddle undergoes a continuous quantum phase transition at intermediate couplings. At the transition the matrix eigenvalue distribution becomes disconnected. The critical excitations are described by large $N$ matrix quantum mechanics. At the critical point, the low energy excitations are waves propagating in an emergent $1+1$ dimensional spacetime.

\end{abstract}

\pagebreak
\setcounter{page}{1}

\newpage

\tableofcontents

\newpage

\section{Introduction}

The phrase `It from Qubit' expresses the intuition that universal quantum computing, of discrete qubits, is the correct framework for thinking about physical reality (e.g. \cite{dd}). Recent developments on several fronts have emphasized the power of quantum information theoretic ideas in characterizing physical systems. Firstly, phases of matter with the same symmetries can have differing quantum order \cite{Wen:2004ym}, and this order is quantified through the entanglement in the quantum state of the system \cite{kp,lw}. Secondly, the ground state of a quantum system with local dynamics is strongly constrained by a necessary accumulation of short distance entanglement, reflected in an area law in the entanglement entropy \cite{a1,a2, a3}. Thirdly, the Ryu-Takayanagi formula \cite{Ryu:2006bv} suggests that the emergence of spacetime itself requires a large amount of microscopic entanglement in `stringy' degrees of freedom that provide the `architecture of spacetime' \cite{Susskind:1994sm, Fiola:1994ir, Bianchi:2012ev,Faulkner:2013ana}. These three ideas are intimately related: Emergent spacetime requires an emergent locality that will necessarily be reflected in the entanglement structure (`quantum order') of the underlying microscopic quantum state \cite{Hartnoll:2015fca}.

In this paper we give a completely explicit realization of the emergence of two dimensional spacetime (`It') from a system of a large but finite number of interacting qubits. The logic we follow contains two steps, as illustrated in figure \ref{fig:twostep} below.
\begin{figure}[h]
\centering
\includegraphics[height = 0.21\textheight]{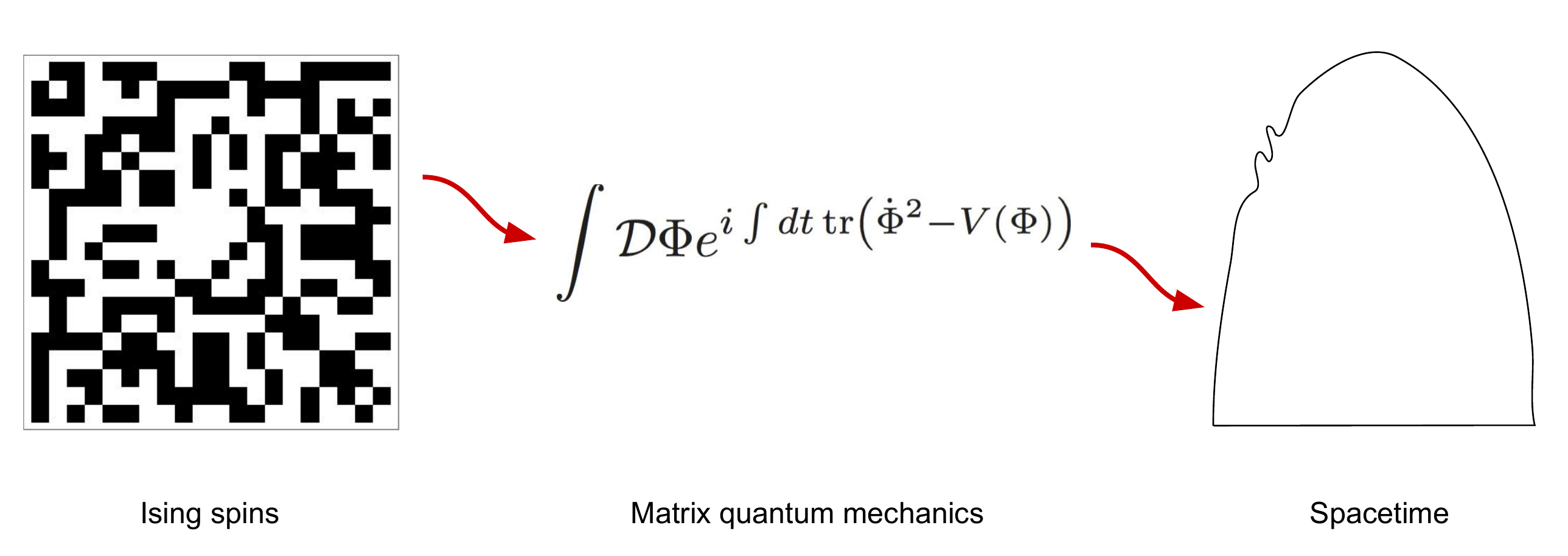}
\caption{\label{fig:twostep} A certain nonlocal transverse field Ising model will be shown to admit a continuum limit given by a matrix quantum mechanics for a single real, symmetric large $N$ matrix. Low energy excitations of this matrix quantum mechanics are described by collective excitations of an emergent semiclassical eigenvalue distribution.}
\end{figure}
The starting point is a particular nonlocal transverse-field Ising model with order $N^2$ spins. Conventional local transverse-field Ising models admit continuum limits close to quantum critical points in which they are described by quantum field theories \cite{sbook}. Instead of a quantum field theory, we wish to engineer a large $N$ matrix quantum mechanics. To achieve this, we show that a particular large $N$ saddle our model undergoes a continuous quantum phase transition. At the critical point, a continuum limit is possible and the critical excitations will be shown to be described by large $N$ matrix quantum mechanics for a single real symmetric matrix. This generalizes a previously established connection between nonlocal classical spin systems and matrix models \cite{parisi, Anninos:2014ffa} to a fully fledged quantum mechanical correspondence.

We proceed to solve the matrix quantum mechanics using standard techniques \cite{Brezin:1977sv, Jevicki:1979mb, Andric:1982jk, Gomis:2003vi}. At the quantum critical point, non-singlet modes are decoupled and we can focus on the gapless singlet (eigenvalue) sector \cite{Gross:1990md, Boulatov:1991xz, Maldacena:2005hi}. The low energy dynamics is then described by the propagation of a collective field in an emergent 1+1 dimensional spacetime. This spacetime is closely related to that of the target space dynamics of the `tachyon' field in two dimensional string theory \cite{Klebanov:1991qa, Ginsparg:1993is, Polchinski:1994mb, Gomis:2003vi}.

The model we study involves a symmetric $N\times N$ matrix worth of spins, with Hamiltonian
\be\label{eq:H0}
H = - h \sum_{A,B=1}^N S^{\,x}_{AB} + \frac{v_4}{N} \sum_{A,B,C,D=1}^N S_{AB}^{z} S_{BC}^{z} S_{CD}^{z} S_{DA}^{z} \,.
\ee
We consider the `antiferromagnetic' case with the coupling $v_4>0$. The classical limit of this model, $v_4 \to \infty$, was previously studied at temperatures $T>0$ in \cite{parisi}. We instead work in the quantum $T = 0$ regime with the new transverse field term in (\ref{eq:H0}). The model is invariant under an $O(N,\Z)$ symmetry (the hyperoctahedral group). 
The classical quartic Ising spin interaction in (\ref{eq:H0}) favors certain `antiferromagnetic crystalline' ordered states \cite{parisi}. The symmetries of the model are restored by either thermal fluctuations or quantum fluctuations induced by the transverse field term in the Hamiltonian (\ref{eq:H0}).

At any nonzero temperature or transverse field, the symmetry breaking dynamics in the model is subtle. The classical ($v_4 = \infty$) model is known to exhibit a structural glassy phase below a critical temperature (shown schematically in figure \ref{fig:phaseplot} below). Numerical study will presumably be needed to see if a quantum spin glass phase survives at $T=0$ and finite $v_4$. The focus of this paper, however, will not be on symmetry breaking. Instead, we will describe the physics of a specific large $N$ saddle point of the partition function. This saddle is singled out by the fact that the $O(N,\Z)$ symmetry is enhanced to $O(N)$. The saddle point will be seen to capture the correct large $N$ physics at small and large coupling. We have not proven that it remains the dominant saddle at all couplings, in particular at the critical coupling where a quantum phase transition occurs.
From the perspective of realizing an explicit emergent spacetime, a large $N$ metastable saddle would be a good enough starting point; after all, our entire universe is likely such a metastable saddle \cite{Anninos:2012qw}.

We will argue that the $O(N)$ invariant `matrix saddle' undergoes a continuous topological quantum phase transition at some $v_4 = v_4^\text{QCP} \sim h$, in which the large $N$ ground state eigenvalue distribution becomes disconnected.\footnote{A similar phase transition occurs in Ising models on planar random graphs \cite{Kazakov:1986hu} -- at present we do not see a direct relation between the random graph model and ours.} Our results combined with those in \cite{parisi} lead to the phase diagram of figure \ref{fig:phaseplot} below for the `matrix saddle'. The connectivity of the eigenvalue distribution is shown in blue. The red line connects our quantum phase transition to the finite temperature transition present in the classical model \cite{parisi}. 

\begin{figure}[h]
\centering
\includegraphics[height = 0.35\textheight]{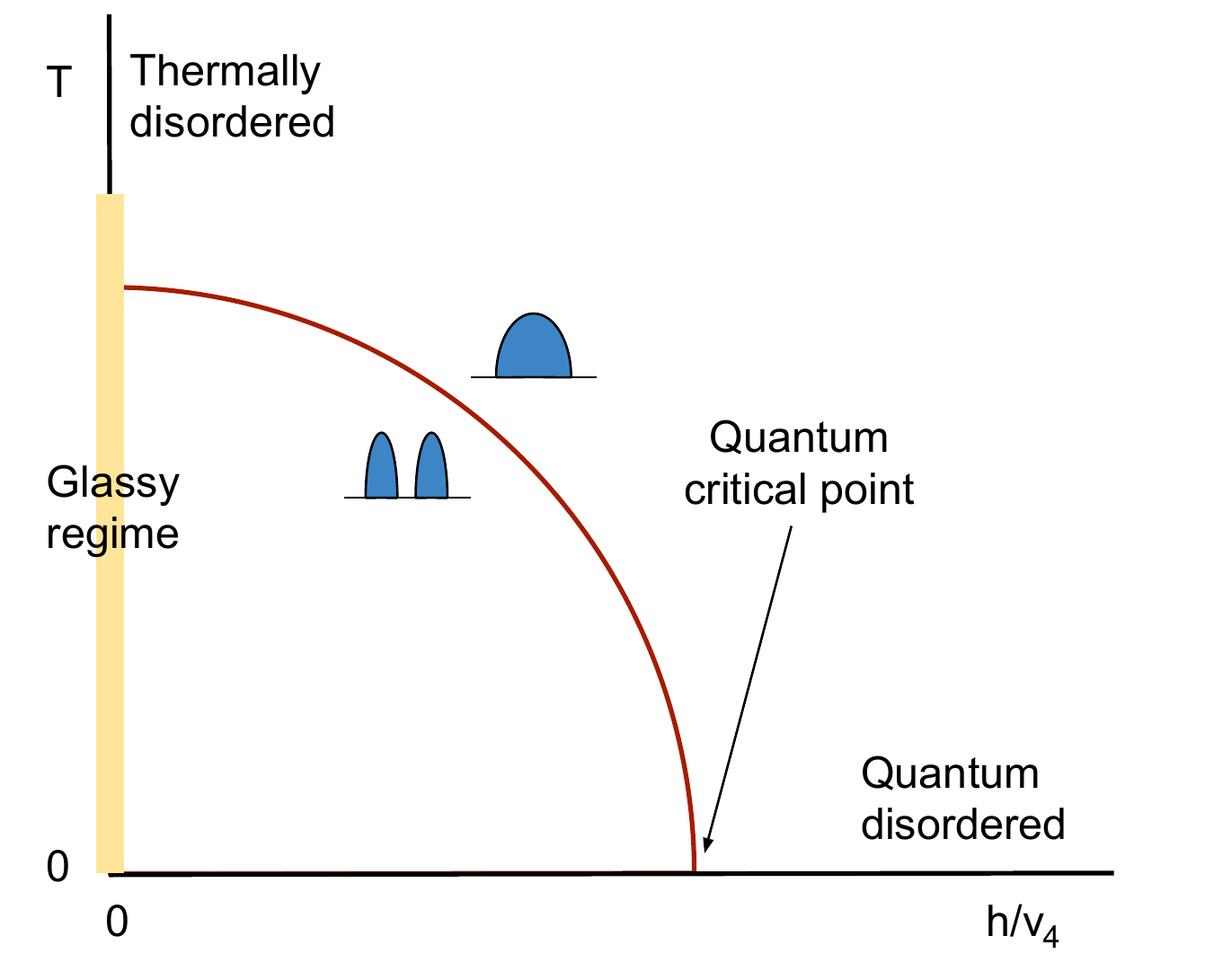}
\caption{\label{fig:phaseplot} Phase diagram of the large $N$ `matrix saddle' of the transverse-field Ising model (\ref{eq:H0}) as a function of temperature $T$ and coupling $v_4$. At the quantum critical point the eigenvalue distribution becomes disconnected. The critical dynamics is described by matrix quantum mechanics.}
\end{figure}

At the quantum critical point shown in figure \ref{fig:phaseplot}, the low energy physics can be mapped to the dynamics of a local field $\phi(t,q)$ in an emergent 1+1 dimensional spacetime, obeying
\be\label{eq:locality}
\ddot \phi - c^2 \pa^2_q \phi = 0 \,.
\ee
The waves propagate in a box whose length $L$ diverges logarithmically as $v_4 \to v_4^\text{QCP}$, leading to the existence of gapless critical excitations. The speed $c$ and spatial coordinate $q$ will be defined below. The emergent local dynamics (\ref{eq:locality}) of this low energy singlet sector (of single-matrix quantum mechanics) from a spin system is a small step towards obtaining truly interesting spacetime physics from qubits. In particular, the low energy singlet sector does not have enough degrees of freedom to describe black holes or other manifestations of stringy entanglement, c.f. \cite{jm}. It does, however, exhibit emergent local dynamics in an emergent spacetime, with the associated accumulation of short distance entanglement \cite{Hartnoll:2015fca, Das:1995vj,Das:1995jw}. It thereby provides the simplest realization of emergent spacetime dynamics.

The remainder of the paper proceeds as follows. In section \ref{sec:model} we introduce the nonlocal transverse-field Ising model in more detail. Section \ref{sec:mapto} shows that the Ising model partition function admits a large $N$ `matrix saddle' with an emergent $O(N)$ symmetry. In section \ref{sec:collectiveH} we derive the large $N$ collective field Hamiltonian describing the eigenvalue dynamics of the corresponding matrix quantum mechanics. The ground state of the collective field Hamiltonian is shown to have a connected eigenvalue distribution at small $v_4$ and a disconnected eigenvalue distribution at large $v_4$, in section \ref{sec:gsm}. In both of these limits, the matrix saddle is shown to capture the ground state of the spins. This section also describes the continuous quantum phase transition separating these two regimes. Section \ref{sec:excitations} characterizes the low energy excitations about the matrix quantum mechanics ground state. At the quantum critical point, gapless excitations propagating locally in an emergent 1+1 dimensional spacetime are found. The conclusion in section \ref{sec:discuss} touches on connections with discrete gauge theories, the possibility of realizing fast scrambling dynamics without quenched disorder, fermionic realizations of qubits, and two dimensional string theory.

\section{A nonlocal transverse-field Ising model}
\label{sec:model}

The model will be built out of a symmetric matrix worth of spin-half operators.
That is
\be\label{eq:S}
S^{\,i}_{AB} = S^{\,i}_{BA} \,, \qquad A,B = 1 \ldots N \,,\qquad i = x,y,z \,.
\ee
We will normalize the operators so that each $S^{\,i}_{AB}$ has eigenvalues $\pm 1$.
The objective is to write down a Hamiltonian for these spins such that the dynamics in the
large $N$ limit can be represented by a bosonic matrix quantum mechanics of the type first
solved in \cite{Brezin:1977sv}. The condition that the matrix be symmetric may not be essential, but
is technically convenient. Let us emphasize that while both e.g. $S^i_{12}$ and $S^i_{21}$ will
appear in the Hamiltonian, they are the same operator and both act on the same spin $|12\rangle$.

We will study Hamiltonians of the following form
\be\label{eq:H}
H = - h \sum_{A,B} S^{\,x}_{AB} + \tr \, V(S^{z}) \,.
\ee
Here and elsewhere, $\tr$ refers to a trace over the $A,B$ indices of (\ref{eq:S}). The potential $V$ in the Hamiltonian is a function of the matrix of $S^{z}$ operators. That is, the terms in $V$ are given by matrix multiplication of the $S^{z}_{AB}$. $V$ is therefore a nonlocal interaction. These interaction terms all commute with each other.
Where we wish to give concrete results, we will work with a microscopic quartic $V$ as in (\ref{eq:H0}) above. However, the universal properties of the critical point we will describe do not depend on the potential, so long as the matrix multiplication structure is present. The statistical physics of some models in this class was studied in \cite{parisi,Anninos:2014ffa}. Here we are interested in the full-blown quantum dynamics induced by the first, transverse field, term in the Hamiltonian (\ref{eq:H}). The role of the transverse field term is to `quantum disorder' the ground state created by the $\tr \, V(S^{z})$ interactions.

The Hamiltonian (\ref{eq:H}) enjoys a $\Z_2^N$ symmetry. The symmetries are generated by acting on the state with $Q_i \equiv \prod_{j=1}^N S_{ij}^{\,x} S_{ji}^{\,x}$, for each $i = 1, \cdots, N$. That is, the spin is flipped along an entire row and corresponding column, with the spin on the diagonal being flipped twice. There is furthermore an $S_N$ symmetry given by permuting rows and simultaneously the corresponding columns. The full symmetry group generated by these symmetries is the hyperoctahedral group $O(N,\Z)$. This can be seen as follows. The columns of matrices in $O(N,\Z)$ are orthonormal with integer components. The components $n_i$ of a given column therefore satisfy $\sum_i n_i^2 = 1$. This requires one of the $n_i$ to equal $\pm 1$ and the remainder to be zero. This holds for all columns, which must furthermore be orthogonal to each other. It is clear that such a matrix then describes `signed permutations', which are precisely the symmetries generated by the semi-direct product of $S_N$ and $\Z_2^N$. In disordered phases of the model, these symmetries are unbroken. We can emphasize that there is no $O(N)$ symmetry at this point (the easiest way to see this is that a rotation of a matrix with $\pm 1$ entries is generically not another matrix with $\pm 1$ entries). However the group $O(N,\Z)$ is manifestly a subgroup of $O(N)$.

The dynamics is conveniently encoded in the quantum partition function at inverse temperature $\beta$. Thus we write ($\Tr$ is a trace over the Hilbert space)
\be
Z =  \Tr \, e^{- \beta H} \,.
\ee
We now carry out a series of standard steps in moving towards a path integral description of the partition function. That is, we introduce a large number $M$ of resolutions of the identity in terms of the simultaneous eigenstates $S^{z}_{AB} | \sigma \rangle = \sigma^{AB} | \sigma \rangle$ and perform a Suzuki-Trotter decomposition. Thus
\bea
Z & = &  \sum_{\sigma_1 \in \{\pm 1\}^{N^2}} \cdots \sum_{\sigma_M \in \{\pm 1\}^{N^2}}  \; \prod_{m=1}^M \langle \sigma_m | e^{- \vep H}| \sigma_{m+1} \rangle \,, \qquad \vep \equiv \frac{\beta}{M} \ll 1  \label{eq:partition} \\
& \approx &  \sum_{\sigma_1 \in \{\pm 1\}^{N^2}} \cdots \sum_{\sigma_M \in \{\pm 1\}^{N^2}}  \;  \prod_{m=1}^M e^{- \vep \,  \tr V_o(\sigma_m)} \prod_{A,B} \langle \sigma_m | 1 + \vep \, h \, S^{\,x}_{AB}| \sigma_{m+1} \rangle \nonumber \\
& = & \sum_{\sigma_1 \in \{\pm 1\}^{N^2}} \cdots \sum_{\sigma_M \in \{\pm 1\}^{N^2}}  \; 
\exp\left\{ - \vep \sum_{m=1}^M \left( \tr \, V_o(\sigma_m) + \frac{\widetilde K}{2} \sum_{A,B} \frac{\left(\sigma^{AB}_m - \sigma_{m+1}^{AB} \right)^2}{\vep^2}\right)\right\} \,. \nonumber
\eea
In these sums there are $N^2$ independent spins only before the symmetry constraint (\ref{eq:S}) is imposed.
The expansion of the exponent in the second line requires $\vep h \ll 1$.
In the final line we have set $\widetilde K = \vep/2\times \log (\vep h)$. The quantity $\vep \equiv \beta/M$ itself is an auxiliary variable, not part of the microscopic model.

\section{Map to a constrained matrix quantum mechanics}
\label{sec:mapto}

In this section we map the Ising spin partition function (\ref{eq:partition}) onto that of a matrix $\Phi$ of continuous bosonic fields. Some of the steps we go through are familiar from the well-understood mapping of spins onto bosons  in local Ising models. However, there are some particularities in our case because in quantum mechanics (with no spatial dimension, i.e. as opposed to quantum field theory), higher order interactions such as $\Phi^{2n}$ are increasingly relevant rather than irrelevant. Furthermore, our continuous phase transition will not manifestly be a symmetry breaking transition. For these reasons, we go through the steps carefully.

The sums over spin states $\sigma^{AB}_m$ in the partition function (\ref{eq:partition}) can be exchanged for integrals over symmetric matrices of bosons $\Phi^{AB}_m$ as follows:
\bea
\sum_{\sigma_1 \in \{\pm 1\}^{N^2}} \cdots \sum_{\sigma_M \in \{\pm 1\}^{N^2}} F(\sigma) & = & \int  \prod_{m,A,B} d\Phi_m^{AB} \delta\left( \left(\Phi_m^{AB}\right)^2 - 1 \right) F(\Phi) \\
& = & \int  \prod_{m,A,B} d\Phi_m^{AB} d\mu_m^{AB} e^{i \mu_m^{AB} \left( \left(\Phi_m^{AB}\right)^2 - 1 \right)} F(\Phi) \,, 
\eea
for any function $F$. In both of these steps, we have dropped numerical prefactors ($2$'s and $\pi$'s) that will only contribute to an unimportant overall normalization of the partition function. In the second step we furthermore introduced Lagrange multipliers $\mu_m^{AB}$ to impose the constraints. While the two steps above are rather trivial, this reformulation will be especially powerful in the large $N$ limit.

The continuum limit corresponds to taking the $\vep \to 0$ limit of the imaginary time discretization, so that $\Phi^{AB}_m \to \Phi^{AB}(\tau)$. Such continuum limits are subtle in spin systems. The $\vep$ in $\widetilde K$ in (\ref{eq:partition}) means that as $\vep \to 0$ discontinuous paths contribute to the path integral. With discontinuous paths, one is not even guaranteed that the Riemann integral exists. In the expectation (and we will later see this explicitly) that the large $N$ limit will be powerful enough to favor sufficiently well-behaved paths, we put this concern aside and, as is familiar, the first term in the final exponent of (\ref{eq:partition}) becomes
\be
\vep \sum \tr \, V(\Phi_m) \to \int d\tau \, \tr \, V(\Phi) \,.
\ee
The presence of jagged and even discontinuous paths at any fixed nonzero $\vep$ is more serious for the remaining finite difference term in the final exponent of (\ref{eq:partition}). In general, we must allow higher derivative terms in the replacement:
\be
\frac{\widetilde K}{2} \frac{\left(\Phi_m - \Phi_{m+1} \right)^2}{\vep^2} \; \to \; \frac{K}{2} \left( \frac{d\Phi}{d\tau} \right)^2 + \frac{K'}{2}  \frac{1}{h^2} \left( \frac{d^2\Phi}{d\tau^2} \right)^2  + \cdots \,. \label{eq:ddt}
\ee
Higher order time derivatives come with inverse powers of the microscopic energy scale $h$. Away from the quantum critical point, on the disordered (small $v_4$) side, typical excitations have energies of order $h$, and the higher derivative terms are generically important. The coefficients $K,K'
,\ldots$ are to be determined in the spirit of effective field theory, by matching with the microscopics. This expansion can be truncated to the leading power for excitations with energies $\Delta E \ll h$. Such states will be seen to exist at the gapless quantum critical point described below. Therefore, it is only close to the quantum critical point where the continuum limit theory, keeping the lowest order derivative term in (\ref{eq:ddt}), can be expected to correctly capture the low energy physics. We will see that the leading order kinetic term is also sufficient in the limiting cases of weak $v_4$ coupling, to three but not four orders in perturbation theory for the ground state energy, and to leading order at strong $v_4$ coupling.

Keeping the leading time derivative term in (\ref{eq:ddt}), with the above caveats in mind, the continuum limit of the partition function (\ref{eq:partition}) becomes
\be
Z = \int {\mathcal D} \Phi {\mathcal D} \mu \exp\left\{ - \int_0^{\beta} d\tau \left[ \tr \left(\frac{K}{2} \left( \frac{d\Phi}{d\tau} \right)^2 + V(\Phi) \right) + i \sum_{AB} \mu_{AB} \left( \left(\Phi^{AB}\right)^2 - 1 \right) \right] \right\} \,. \label{eq:Z1}
\ee
The partition function is now a quantum mechanical path integral over two symmetric matrices $\Phi$ and $\mu$. It is not yet a matrix quantum mechanics, however, as the final term in (\ref{eq:Z1}) is not $O(N)$ invariant (i.e. it does not have the form of a matrix multiplication). This is expected given that the steps so far have been exact (up to subtleties with taking the continuum limit), and the original model was not $O(N)$ invariant.

A genuine matrix quantum mechanics is obtained as follows. Here we take inspiration from \cite{parisi}, as we discuss in more detail below. First imagine integrating out the matrix $\Phi$ in (\ref{eq:Z1}) to obtain an effective theory for the matrix of Lagrange multipliers $\mu$. This integral inherits an $S_N$ symmetry from the full partition function, corresponding to permuting rows and columns of $\mu$. It is consistent to look for large $N$ saddle points that are invariant under this symmetry. These are matrices where all off-diagonal terms in the matrix of Lagrange multipliers are equal and all diagonal terms are also equal: $\mu_{AB} = - i \mu - i \tilde \mu \delta_{AB}$. In such a saddle the constraint becomes
\be\label{eq:relax}
i \sum_{AB} \mu_{AB} \left( \left(\Phi^{AB}\right)^2 - 1 \right) \to \mu \, \tr \left(\Phi^2 - N \right) + \tilde \mu \sum_A \left(\Phi_{AA}^2 - 1 \right) \,.
\ee
The second of these terms is subleading at large $N$ and can be dropped to leading order (that is to say, the spins along the diagonal of the matrix correspond to $N$ out of order $N^2$ variables, and hence can be neglected -- we will see some concrete evidence for this later). The partition function then becomes
\be
Z_\text{matrix} = \int {\mathcal D} \Phi {\mathcal D} \mu \exp\left\{ - \int_0^{\beta} d\tau \, \tr \left(\frac{K}{2} \left( \frac{d\Phi}{d\tau} \right)^2 + \mu \Phi^2 + V(\Phi) - \mu N \right)\right\} \,. \label{eq:Z3}
\ee
Here $\mu$ is a single field, not a matrix. We will refer to large $N$ saddles captured by (\ref{eq:Z3}) as `matrix saddles'. In these saddles, the $O(N,\Z)$ symmetry of the original Hamiltonian (\ref{eq:H}) has been enhanced to $O(N)$.

The `softening' of spins at large $N$ by relaxing the individual normalization constraints has a long history, going back to the classical `spherical model' \cite{Berlin:1952zz, Stanley:1968gx}. These ideas were applied to the classical version of our nonlocal Ising model by \cite{parisi}, who noted the existence of the `matrix saddle' above.\footnote{
\cite{parisi} also considered an intermediate case where the order $N^2$ constraints are relaxed to $N$ constraints (rather than one constraint). The quartic `matrix multiplication' Ising spin interaction of (\ref{eq:H0}) is written as a quartic interaction of $N$, $N$-dimensional `rotors', $\vec n_A$:
$\sum_{A,B,C,D} S_{AB}^{z} S_{BC}^{z} S_{CD}^{z} S_{DA}^{z} = \sum_{A,C} \left(\vec n_A \cdot \vec n_C\right)^2.$  Each rotor is then normalized as $|\vec n_A|^2 = N$, but the components are unconstrained. Our model may correspondingly be related to a large $N$ quantum rotor model (cf. \cite{sbook}) with a quartic interaction.}
However, in the more established cases, interactions are local and hence the softening of spins is an intuitive process that occurs during a spin-blocking type renormalization group flow. As spins are locally grouped together, the range of values the effective spin can take becomes less constrained. It is unclear that this intuition holds in nonlocal models. However, the numerical Monte-Carlo results in \cite{parisi} show that indeed, outside of the low temperature glassy regime, the matrix saddle correctly describes the classical spin system at all temperatures. The evidence for the dominance of the matrix saddle in our quantum case will be restricted to perturbation theory at weak and strong coupling. As noted in the introduction, a potentially metastable large $N$ saddle point is good enough for our purposes of realizing an emergent spacetime. We therefore proceed to solve the large $N$ constrained matrix quantum mechanics described by (\ref{eq:Z3}).

\section{Collective field Hamiltonian}
\label{sec:collectiveH}

To obtain the ground state wavefunction, we first need the Schr\"odinger equation corresponding to the partition function (\ref{eq:Z3}). This leads us to an exercise in the quantization of constrained systems. The Lagrangian is
\be\label{eq:LLt}
L =  \tr \left(\frac{K}{2} \dot \Phi^T \dot \Phi - V(\Phi) - \mu \left( \Phi^T \Phi - N \right)  - \nu \left(\Phi^T - \Phi \right) \right) \,.
\ee
We have introduced a new matrix $\nu_{AB}$ of Lagrange multipliers in order to impose the constraint that $\Phi$ be symmetric. There will be three steps in this section. Firstly we obtain the quantum mechanical Hamiltonian and constraints following from (\ref{eq:LLt}). This will be equations (\ref{eq:H1}) and (\ref{eq:C1}) below. Secondly we diagonalize the matrix $\Phi$ and obtain the eigenvalue Hamiltonian (\ref{eq:Hlam}). Thirdly, we change variables to a collective field and obtain the final collective field Hamiltonian (\ref{eq:col2}) and constraints (\ref{eq:twoc}). The advantage of this last formulation is that the large $N$ limit can explicitly be treated in the saddle point approximation. I.e. the collective field is the `master field' \cite{witten}.

\subsection{Matrix quantum mechanics Hamiltonian}

Standard manipulations starting from (\ref{eq:LLt}) -- see Appendix \ref{sec:dirac} --  lead to the Hamiltonian
\be\label{eq:Hff}
H = \tr \left(\frac{1}{2 K} \Pi^T \Pi + V(\Phi) \right) \,,
\ee
with $\Pi$ the momentum conjugate to $\Phi$, together with the constraints
\bea
\tr \left(\Phi^T \Phi - N\right) = 0 \,, &\quad& \tr (\Phi^T \Pi) = 0, \label{eq:L1} \\
\Phi - \Phi^T= 0 \,,&\quad& \Pi - \Pi^T = 0 \,. \label{eq:L2}
\eea
The first line constrains the components of $\Phi$ to lie on a high dimensional sphere and furthermore to have no momentum perpendicular to the sphere (and hence to remain on the sphere). As usual with a constrained Hamiltonian system, the dynamics is determined by Dirac rather than Poisson brackets. In this case (see Appendix \ref{sec:dirac})
\be
\{\Phi^{AB}, \Pi^{CD} \}_\text{Dirac} =  \frac{1}{2} \left(\delta^{AC} \delta^{BD} + \delta^{AD} \delta^{BC} \right) - \frac{1}{N^2} \Phi^{AB} \Phi^{CD} \,. \label{eq:dirac}
\ee

Upon quantization, the Dirac bracket (\ref{eq:dirac}) becomes the commutator (with an extra factor of $i$, as usual). This means that the momentum operator must be represented as\footnote{In equations (\ref{eq:mom}) and (\ref{eq:H1}), the operators $\Pi^{CD}$ and $H$ are not manifestly Hermitian. Making them explicitly Hermitian (i.e. by adding the hermitian conjugate and dividing by two) simply leads to an overall constant shift in the Hamiltonian (\ref{eq:H1}).}
\be\label{eq:mom}
\Pi^{CD} = - i\, \overline \pa_{CD} + \frac{i}{N^2} \Phi^{CD} \sum_{MN} \Phi^{MN} \overline \pa_{MN} \,.
\ee
We have set $\hbar = 1$ and introduced the symmetric derivative
\be
\overline \pa_{AB} \equiv \frac{1}{2} \left(\frac{\pa}{\pa \Phi^{AB}} + \frac{\pa}{\pa \Phi^{BA}} \right) \,.
\ee
The second constraint in (\ref{eq:L1}), as well as the second constraint in (\ref{eq:L2}) are automatically satisfied once the momentum is given by (\ref{eq:mom}).
The Hamiltonian then becomes, as a differential operator, 
\be
H = - \frac{1}{2 K} \left[ \bar{\partial} \cdot \bar{\partial} -\frac{N^2+N - 4}{2 N^2}  \Phi \cdot \bar{\partial} -\frac{1}{N^2} \left( \Phi  \cdot \bar{\partial} \right)^2  \right] + \tr \,V(\Phi)  \,. \label{eq:H1}
\ee
Here we defined $S \cdot T \equiv \sum_{AB} S_{AB} T_{AB}$.
We wish to find the ground state of this Hamiltonian, which must be solved together with the operator identities
\be\label{eq:C1}
\tr \left( \Phi^T \Phi \right) = N^2 \,, \qquad \Phi - \Phi^T= 0 \,.
\ee
Use of the Dirac bracket has ensured these constraints commute with the Hamiltonian (\ref{eq:H1}).

\subsection{Eigenvalue Hamiltonian}

The symmetric matrix $\Phi$ can be diagonalized using an orthogonal matrix $O$:
\be
\Phi = O^T \Lambda O \,, \qquad \Lambda_{ij} = \lambda_i \delta_{ij} \,.
\ee
The Hamiltonian (\ref{eq:H1}) becomes (some details are given in the Appendix)
\be
H = H_\lambda + H_O\,.
\ee
The eigenvalue part of the Hamiltonian is
\be\label{eq:Hlam}
H_\lambda = -\frac{1}{2K} \left[ \sum_i \frac{1}{\Delta(\lambda)} \partial_i \Delta(\lambda) \partial_i - \frac{N^2+N - 4}{2 N^2} \sum_i \lambda_i \partial_i - \frac{1}{N^2} \left( \sum_i \lambda_i \partial_i \right)^2 \right] + \sum_i V(\lambda_i) \,,
\ee
where
\be
\Delta(\lambda) = \prod_{i<j} (\lambda_i - \lambda_j) \,,
\ee
is the usual Vandermonde measure factor for symmetric matrices. The remaining operator constraint is
\be\label{eq:lambdaconstraint}
\sum_i \lambda_i^2 = N^2 \,.
\ee
It is simple to check that this constraint commutes with the Hamiltonian (\ref{eq:Hlam}), as it should.

The Hamiltonian for the diagonalizing orthogonal matrices is
\be\label{eq:Hoff}
H_O = - \frac{1}{2K} \sum_{i < j} \frac{1}{(\lambda_i - \l_j)^2} \frac{\pa}{\pa \Omega_{ij}} \frac{\pa}{\pa \Omega_{ij}} \, ,
\ee
where
\be\label{eq:Omeg}
d \Omega = dO \cdot O^T \,.
\ee
It is clear that any dependence of the wavefunction on $\Omega_{ij}$ will increase the energy.
The orthogonal matrices $O$ live in a compact space and therefore the ground state wavefunction will simply
be independent of the $\Omega_{ij}$. For a more extended discussion see e.g. \cite{Klebanov:1991qa}. The immediate upshot is
that in discussing the ground state, we can simply ignore $H_O$. It will become important later when we wish to consider
excitations. In order for the eigenvalue dynamics to capture the low energy physics, it will be important that non-singlet modes are sufficiently heavy. This will indeed be the case at the quantum critical point.

\subsection{Collective field Hamiltonian}

Hamiltonians such as (\ref{eq:Hlam}), for the eigenvalues of real symmetric matrices, are Calogero-Moser models. In particular, unlike for Hermitian matrix quantum mechanics, the eigenvalues experience interactions leading to generalized statistics (see e.g. \cite{Gomis:2003vi}). At large $N$, the ground state of this Hamiltonian is most easily characterized using the collective field method \cite{Jevicki:1979mb, Andric:1982jk}. In this approach
the eigenvalue density $\rho(\lambda,t)$ is introduced as
\be
\rho(\lambda,t) = \sum_{i=1}^N \delta\left(\lambda - \lambda_i(t) \right) \,.
\ee
We define the canonical conjugate momenta to be $\pi(\l,t)$, so that
\be
\left[\rho(\l,t), \pi(\l',t)\right] = i \delta(\l-\l') \,.
\ee
The partial derivatives in the Hamiltonian (\ref{eq:Hlam}) can be expressed in terms of $\pi(\lambda) = - i \delta/\delta \rho(\lambda)$ using the chain rule. Thus
\be
\pa_i =-  i \int d \lambda \, \delta'(\lambda - \lambda_i) \, \pi(\lambda) = i \pa_\l \pi(\l) \Big|_{\l = \l_i} \,.
\ee
The large $N$ Hamiltonian then becomes
\bea
H = \int d\lambda \rho(\lambda) \Bigg(\frac{1}{2K} \left[\pa_\lambda \pi(\lambda) \pa_\lambda \pi(\lambda) - \frac{i}{2} \pa_\l^2 \pi(\l) - i \rho_H(\l) \pa_\lambda \pi(\l) \right]+ V(\lambda)  \nonumber \\
 + \frac{i}{4 K} \lambda \pa_\lambda \pi(\lambda) \left[1 + \frac{2 i}{N^2}   \int d\l' \rho(\l') \, \l'  \pa_{\l'} \pi(\lambda') \right] + \frac{i}{2 K N^2} \lambda \pa_\lambda \left[\l \pa_{\l} \pi(\l) \right] \Bigg) \,. \label{eq:col1}
\eea
Here the Hilbert transform
\be\label{eq:rhoH}
\rho_H(\l) = \int d\lambda' \rho(\l') \frac{P}{\l - \l'} \,.
\ee
The factor of $1/2$ in the second term in the first line of (\ref{eq:col1}) is somewhat subtle, and arises together with taking the principal value of the integral in (\ref{eq:rhoH}).

There are now two constraints
\be\label{eq:twoc}
\int d\l \rho(\l) = N \,, \qquad \text{and} \qquad \int d\l \l^2 \rho(\l) = N^2 \,.
\ee
These two constraints require the $N$ scaling
\be\label{eq:bel}
\rho  = \sqrt{N} \, \hat \rho \,, \qquad \l = \sqrt{N} \, \hat \l \,.
\ee
From this scaling we can immediately see that the second term in the first line of (\ref{eq:col1}) is subleading in $N$ compared the third term. The last term in the second line is similarly subleading. We will drop these two terms henceforth. We will see shortly that in the ground state of interest, $\pi = N \hat \pi$. This means that all the remaining terms are of the same order and must be kept. We can also see that for the potential to compete with the other terms it must scale as $V = \hat V \, N/K$.

The Hamiltonian (\ref{eq:col1}) is not manifestly Hermitian because there is a nontrivial measure factor in the wavefunction normalization. This factor can be removed by rescaling the wavefunction, which amounts to shifting the momenta $\pi(\lambda) \to \pi(\lambda) + X(\lambda)$, as explained in \cite{Jevicki:1979mb}. The shift $X$ is chosen to remove the linear-in-momenta terms from (\ref{eq:col1}). Thus we need $X$ to satisfy
\be
2 \, \pa_\l X(\lambda) - \frac{2}{N^2} \int d\l' \rho(\l') \l \l' \, \pa_{\l'} X(\l') = i \rho_H(\l) - \frac{i}{2} \lambda \,.
\ee
We have dropped the subleading in large $N$ terms identified in the sentences below (\ref{eq:bel}).
The general solution to this equation (using both of the constraints in (\ref{eq:twoc})) is
\be
\pa_\l X(\lambda) = c \, \l + \frac{i \rho_H(\l)}{2} \,,
\ee
where $c$ is an arbitrary constant. We only need to find one $X(\l)$ that does the job of making $H$ explicitly Hermitian, and so we can set $c = 0$ without loss of generality (in fact, $c$ can be shown to drop out of the final results in any case). The large $N$ Hamiltonian is now
\bea
H & = & \int d\l \rho(\l)  \Bigg(\frac{1}{2 K} \left[\pa_\lambda \pi(\lambda) \pa_\lambda \pi(\lambda) + \frac{1}{4} \rho_H(\l)^2  \right]+ V(\lambda) \Bigg) \nonumber \\
& & - \; \frac{1}{2 K N^2} \left(\int d\l \r(\l) \l \pa_\l \pi(\l) \right)^2 - \frac{N^2}{32 K} \,. \label{eq:col2}
\eea
We have used the constraints (\ref{eq:twoc}) to simplify these terms. Various terms that arise in commuting $\rho$'s and $\pi$'s are subleading at large $N$. The final collective field Hamiltonian (\ref{eq:col2}) of course commutes with the constraints (\ref{eq:twoc}). From the Hamiltonian (\ref{eq:col2}) we can now characterize the ground state as well as the low energy collective eigenvalue excitations.

\section{Ground state}
\label{sec:gsm}
 
At large $N$, the ground state is found by classically minimizing the Hamiltonian (\ref{eq:col2}) subject to the constraints (\ref{eq:twoc}). This semiclassical approach holds because, in the collective field path integral, typical configurations have action (and energy) of order $N^2$, while there is now a single field degree of freedom $\rho(\l)$.
Classically the momentum vanishes in the ground state, so that $\pi = 0$.  Therefore we must minimize
\be\label{eq:Erho}
E[\rho] = \int d\l \rho(\l) \left( \frac{\pi^2}{24K} \rho(\l)^2 + V(\l) - c_1 - c_2 \l^2 \right) + c_1 N + c_2 N^2 - \frac{N^2}{32 K} \,.
\ee
We introduced Lagrange multipliers $c_1$ and $c_2$ to impose the constraints. We also used the identity (that we learnt from \cite{Andric:1982jk}) that
\be\label{eq:simp}
\int d\l \r(\l) \r_H(\l)^2 = \frac{\pi^2}{3} \int d\l \rho(\l)^3 \,.
\ee
It is trivial now to minimize (\ref{eq:Erho}).

We will specialize at this point to the quartic potential
\be\label{eq:v}
V(\l) = \frac{v_4}{N} \l^4 \,.
\ee
We take $v_4>0$, in order for the minimum of (\ref{eq:Erho}) to be stable. The model (\ref{eq:H}) we are studying then corresponds to a quantum disordering of the quartic nonlocal classical Ising model considered in \cite{parisi}. The coupling $v_4$ is kept fixed in the large $N$ limit. It is also convenient to introduce
\be\label{eq:vhat}
\hat v_4 = K v_4 \,,
\ee
which will play the role of a dimensionless coupling in the continuum theory.

Following the discussion around equation (\ref{eq:ddt}) above --- concerning the continuum limit of the discrete time derivative --- the collective field Hamiltonian (obtained via matrix quantum mechanics) is only guaranteed to correctly describe the low energy excitations at a quantum critical point. Our presentation in the remainder will be as follows. We will find the ground state and the low energy excitations of the collective field Hamiltonian (\ref{eq:col2}) with the potential (\ref{eq:v}) at all couplings. This will allow us to achieve two things. First, we will be able to isolate the singular behavior and critical excitations at the critical point. These are universal singular properties of the critical point and so should correctly capture the critical behavior of the matrix saddle. Second, we will see that the collective field ground state also correctly reproduces that of the spin model at small and large $v_4$. This will be important for us to argue that indeed there is a continuous quantum phase transition at intermediate coupling in the matrix saddle. Finally, we note that the universal eigenvalue dynamics at a continuous quantum critical point is independent of the choice (\ref{eq:v}) of potential.

The distribution that minimizes the energy (\ref{eq:Erho}) can be written
\be\label{eq:dist}
\rho_\star(\l) = \frac{\sqrt{8 N}}{\pi} \theta\left(\l_o^2 - \frac{\l^2}{N}\right) \sqrt{\left(\l_o^2 - \frac{\l^2}{N}\right)\left(a^2 + \hat v_4 \frac{\lambda^2}{N} \right)} \,.
\ee
We have assumed that the parameters are such that the eigenvalue distribution is only a single connected component, with range $\lambda \in [-\sqrt{N}\l_o,\sqrt{N}\l_o]$. This requires the parameter $a^2 > 0$. The parameters $\{\l_o,a\}$ depend on $\{c_1,c_2\}$ in (\ref{eq:Erho}), although we will not need the explicit relation.  We will discuss shortly the phase transition associated with $a^2$ becoming negative, in which the eigenvalue distribution becomes disconnected. The eigenvalue density $\rho$ must of course be real and nonnegative everywhere.

\subsection{Behavior and matching at small $v_4$}
\label{sec:smallv}

The constraints (\ref{eq:twoc}) determine $\l_o$ and $a$ in terms of the couplings. The integrals can be performed explicitly in an expansion at small $\hat v_4$ to give
\bea
a & = & \frac{1}{4 \sqrt{2}}  - 48 \sqrt{2} \hat v_4^2 + \cdots \,,\\
\l_o & = & 2 - 16 \hat v_4 + 1088 \hat v_4^2 + \cdots  \,.
\eea
It is easy to obtain the perturbative solution to high orders. One can simply expand the distribution (\ref{eq:dist}) in small $\hat v_4$ inside the integral. Note that the $\hat v_4 = 0$ distribution is a Wigner semicircle and the integrals that arise are elementary.

An important class of observables are the single trace moments of the eigenvalue distribution, which correspond microscopically to traces of powers of the matrix of spins. In a weak coupling expansion one obtains
\bea
\lefteqn{\frac{1}{N^{n+1}} \lefteqn{\langle \tr \, \Phi^{2n} \rangle = \frac{1}{N^{n+1}} \int d\lambda \rho(\l) \l^{2n} =} } \label{eq:moments}  \\
 && \frac{2 \, \Gamma(2n)}{\Gamma(n) \Gamma(n+2)} \left(1  - \frac{16 n (n-1)}{n+2} \hat v_4 +
\frac{128 n(n-1) (10+8n+n^2)}{(n+2)(n+3)} \hat v_4^2 +
\cdots \right) \,. \nonumber
\eea
These observables directly characterize the full eigenvalue distribution. We have verified that the three terms in equation (\ref{eq:moments}) -- i.e. up to order $v_4^2$ --  are precisely reproduced for all $n$ by explicit microscopic computations in the spin system. The matching with microscopics fixes the coefficient of the matrix saddle kinetic term to be
\be\label{eq:Kh}
K = \frac{1}{16 h} \,.
\ee
The zeroth and first order spin computations of the moments are given in Appendix \ref{sec:matching}. Perturbative computation of the moments of the spin system translate into a combinatorial problem that is solved by the matrix quantum mechanics in (\ref{eq:moments}). This matching is one of our main results, we have put the entire computation in an appendix only because it is somewhat technical.

One can also obtain the ground state energy in a weak coupling expansion. The ground state energy is evaluated from (\ref{eq:Erho}) to be
\be
K \frac{E_0}{N^2} = \Lambda + 2 \hat v_4 -8 \hat v_4^2 + 256 \hat v_4^3 - 14848 \hat v_4^4 + \cdots \,. \label{eq:EGS}
\ee
Here we have included an undetermined overall constant $\Lambda$. This is present because we have not kept track of the overall normalization of the partition function and also we have not worried about operator ordering ambiguities that appear in Dirac quantization. This term aside, 
we have reproduced the remaining terms up to order $v_4^3$ in the ground state energy (\ref{eq:EGS}) from standard quantum mechanical perturbation theory in the microscopic spin system. This matching fixes
\be
\Lambda = - \frac{1}{16} \,.
\ee

Despite the above agreements, we have found that at order $v_4^4$, the ground state energy of the matrix quantum mechanics in (\ref{eq:EGS}) does not match that of the spin system. The spin system answer is instead $-15360 \hat v_4^4$. We have obtained the spin answer both numerically and analytically. Directly related to the mismatch in energy at  order $v_4^4$, the moments computed above will also disagree at order $v_4^3$. These mismatches between bosons and spins are in contrast to high temperature perturbation in the classical model of \cite{parisi}, which we have verified agrees through to fourth order and probably to all orders.

The crucial difference between the quantum and classical models is the need in the quantum model to take a continuum limit in time. As discussed around equation (\ref{eq:ddt}) above, this limit is only justified close to a quantum critical point. The simplest interpretation of the mismatch, then, is that beyond the first few orders in perturbation theory in $v_4$, one needs to deal with the discrete time derivative in the bosonic description. An alternative possibility is that the continuum limit is not the source of disagreement, but that the matrix saddle is not the dominant saddle in general, yet happens to coincide with the dominant saddle to low orders in perturbation theory. In either case, there is an important positive outcome from the matching to low orders in perturbation theory. We learn that the leading time derivative term in (\ref{eq:ddt}) is in fact sufficient to describe the matrix saddle to these low orders. Knowing the behavior of the matrix saddle at weak coupling will be important in section \ref{sec:exist} below to argue that the matrix saddle undergoes a quantum phase transition at intermediate couplings.

In obtaining the ground state energy of the spin system to fourth order in perturbation theory, we have found that virtual states involving the diagonal spins never contribute at leading order in $N$. This is consistent with our expectation above that the effects of these $N$ out of $N^2$ degrees of freedom should be subleading at large $N$.

\subsection{Phase transition to a disconnected eigenvalue distribution}
\label{sec:transition}

Moving to larger values of the coupling $\hat v_4$, the most important phenomenon that occurs is a topological phase transition in the eigenvalue distribution. Namely, the eigenvalue distribution becomes disconnected when $a=0$ in the solution (\ref{eq:dist}). Inserting the distribution with $a=0$ into the
constraints (\ref{eq:twoc}) above, the critical coupling is found to be
\be\label{eq:vcc}
\hat v_4 = \hat v_4^\text{QCP} = \frac{9 \pi^2}{500} \approx 0.1776 \,.
\ee
At the critical point the width of the distribution is
\be\label{eq:lcc}
\l^\text{QCP}_o = \sqrt{\frac{5}{2}} \,.
\ee

The singular behavior of the eigenvalue distribution at the critical point leads to a weak non-analyticity in the ground state energy at $\hat v_4 = \hat v_4^\text{QCP}$. The integrals appearing in the constraints (\ref{eq:twoc}) can be evaluated analytically in terms of elliptic functions. Solving the constraints in an expansion about $\hat v_4^\text{QCP} - \hat v_4$ one finds that the leading non-analyticity causes a divergence in the third derivative of the ground state energy as $\hat v_4 \to \hat v_4^\text{QCP}$:
\be
\frac{K}{N^2} \frac{d^3}{d \hat v_4^3} E_0 = \frac{500}{27 \pi^2} \, \frac{1}{\left(\hat v_4^\text{QCP} - \hat v_4\right)} \, \frac{1}{\log^2\left(\hat v_4^\text{QCP} - \hat v_4\right)} + \cdots \,. \label{eq:Esing}
\ee
A third order continuous transition is characteristic of topological changes in eigenvalue distributions \cite{Gross:1980he, Wadia:1980cp}. The divergence above describes the approach to the critical point from the connected side. The singular contribution comes from eigenvalues close to the disconnection point, where there is a divergent density of states. This is why the critical properties are determined universally, independent of the form of the external potential (\ref{eq:v}) experienced by the eigenvalues. As in the emergence of two dimensional string theory from a discretized worldsheet \cite{Klebanov:1991qa}, it is this singular contribution that truly captures the emergence of local spacetime excitations. We will recall in the following section that non-singlet (`off-diagonal') excitations are only parametrically gapped close to the singular region.

\begin{figure}[h]
\centering
\includegraphics[height = 0.36\textheight]{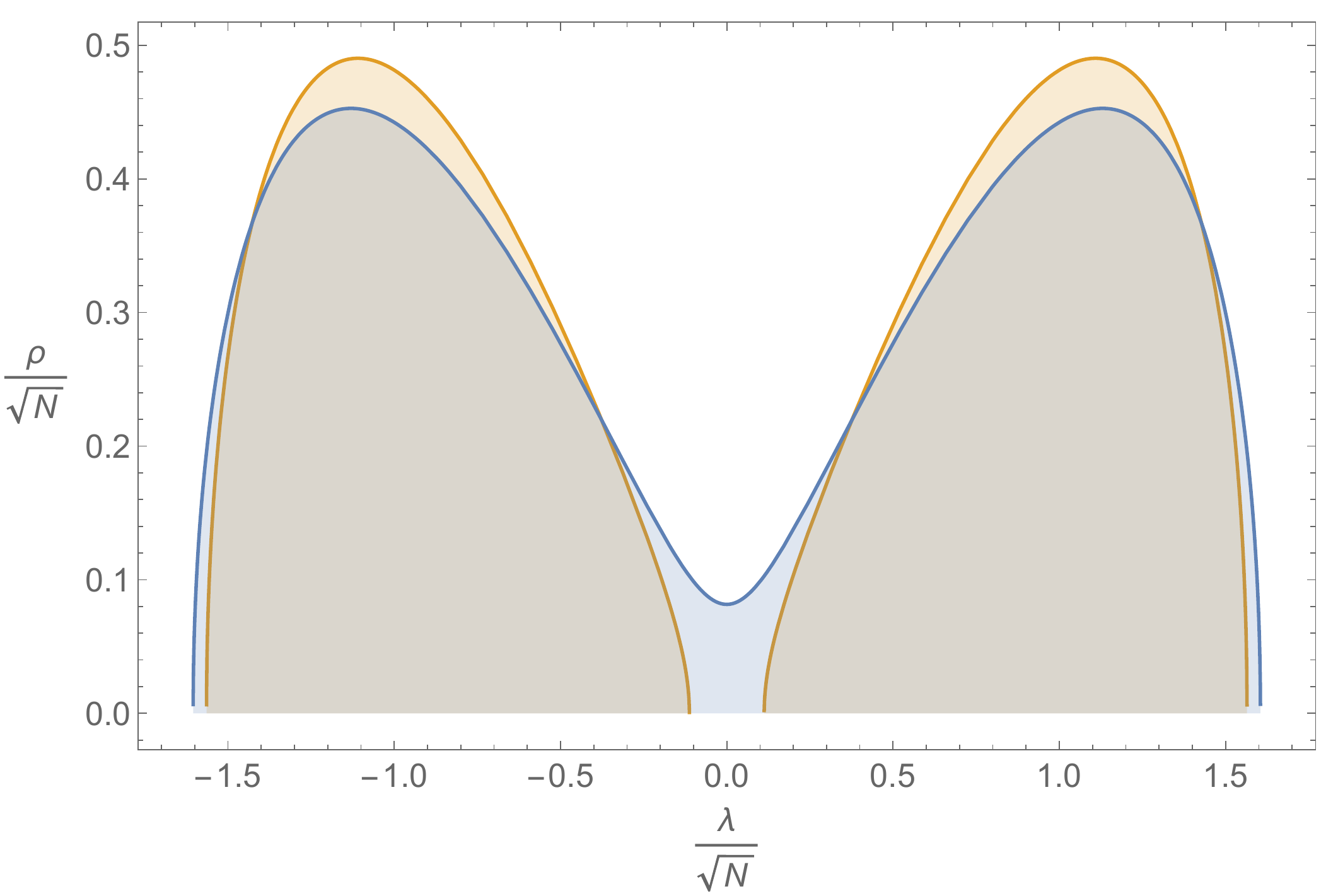}
\caption{\label{fig:rhoplot} Illustrative eigenvalue distributions on each side of the transition. The connected distribution shown has $\hat v_4 = 0.15$, while the disconnected distribution has $\hat v_4 = 0.2$. The critical coupling is $\hat v_4^\text{QCP} \approx 0.177$.}
\end{figure}

Past the critical value of $\hat v_4$, the eigenvalue distribution is disconnected. Thus the solution that minimizes (\ref{eq:Erho}) takes the form
\be\label{eq:dist2}
\rho_\star(\l) = \frac{\sqrt{8 N}}{\pi} \theta\left(\left[\l_+^2 - \frac{\l^2}{N}\right]\left[\frac{\l^2}{N} - \l_-^2\right]\right) \sqrt{\left(\frac{\l^2}{N} - \l_-^2\right)\left(\l_+^2 - \frac{\l^2}{N}\right) \hat v_4} \,.
\ee
The support of the distribution has two components, between $[\sqrt{N}\l_-,\sqrt{N}\l_+]$ and between $[-\sqrt{N}\l_+,-\sqrt{N}\l_-]$, with $0 < \l_- < \l_+$. The constraints may be solved and the energy evaluated in a similar way to what was done in section \ref{sec:gsm}. Illustrative eigenvalue distributions just above and below the critical coupling are shown in figure \ref{fig:rhoplot}.

\subsection{Behavior and matching at large $v_4$}
\label{sec:largev}

Using the disconnected solution (\ref{eq:dist2}) and solving the constraints, one finds that
at large $\hat v_4 \to \infty$ the support of the eigenvalue distribution tends towards two narrow strips with
\be\label{eq:largevL}
\l_- = 1 -  \frac{1}{2 (2 \hat v_4)^{1/4}} + \cdots \,, \qquad \l_+ = 1 + \frac{1}{2 (2 \hat v_4)^{1/4}}  + \cdots  \,.
\ee
There are no quantum fluctuations of the spins in this limit, as the eigenvalue distribution tends towards the sum of delta functions
\be\label{eq:delta}
\rho(\l) = \frac{N}{2} \left( \delta(\lambda + \sqrt{N}) + \delta(\l - \sqrt{N})  \right)\,.
\ee
The moments of the distribution are therefore all given by
\be\label{eq:largemoments}
\frac{1}{N^{n+1}} \langle \tr \, \Phi^{2n} \rangle = \frac{1}{N^{n+1}} \int d\lambda \rho(\l) \l^{2n} = 1 \,.
\ee

The ground state energy as $v_4 \to \infty$ tends to
\be\label{eq:largevE}
\frac{K}{N^2} E_0 = \hat v_4  +  \cdots \,, \quad \Rightarrow \quad \frac{1}{N^2} E_0 = v_4 +  \cdots \,.
\ee
Indeed, the energy scale $K$ associated with quantum fluctuations has dropped out of this formula at leading order, as we should expect. The leading order (linear in $v_4$ term) here exactly reproduces the ground state energy of the matrix phase of the classical spin model found in \cite{parisi}. Note that the energies have been shifted by one in figure 1 of \cite{parisi}. This provides another check on our computations. More importantly however, we now describe how both the ground state energy (\ref{eq:largevE}) and moments (\ref{eq:largemoments}) agree with those of the ground state of the $v_4 \to \infty$ microscopic spin system.

It is convenient to discuss the classical $v_4 \to \infty$ limit by setting $h=0$. The ground states of this model will be some specific matrix configurations $|\text{ordered}\rangle$ of classical Ising spins. Individually these will all necessarily break the symmetries of the model. The energy of these ordered states must be of the form
\be
\frac{1}{N^2} E_\text{ord} = \a \, v_4 \,,
\ee
for some constant $\a$. Determining $\a$ turns out to be a little subtle.

For generic large but finite $N$, the classical ($h=0$) model exhibits glassy physics and so Monte-Carlo computations are unable to find the true ground state. We have performed (classical) Monte-Carlo simulations for various values of $N$ between 50 and 100 and found that states always exist with $\alpha \approx 1.14$ or slightly lower. This is consistent with the energy of glassy states reported in \cite{parisi}.

However, it was noted in \cite{parisi} -- the discussion in section 8 of \cite{parisi2} is also relevant --  that for the special values of $N = 2^k$, with $k$ integer, the exact ground states can be found and these have $\alpha = 1$. These states are then in agreement with the matrix quantum mechanics result (\ref{eq:largevE}). In these states the matrix of Ising spins (taken to be valued in $\pm1)$ is such that all the rows are mutually orthogonal. Symmetric orthogonal matrices are easily constructed iteratively as follows. Firstly for $N=2$, take
\be
M_2 =
\left(
\begin{array}{cc}
1 & 1 \\
1 & -1
\end{array}
\right) \,.
\ee
For general $N = 2^k$ one then has
\be
M_{2^k} =
\left(
\begin{array}{cc}
M_{2^{k-1}} & M_{2^{k-1}} \\
M_{2^{k-1}} & -M_{2^{k-1}}
\end{array}
\right) \,.
\ee
These matrices have all eigenvalues equal to $\pm \sqrt{N}$, also in agreement with the matrices found in the $v_4 \to \infty$ limit of the matrix quantum mechanics described by (\ref{eq:delta}). It immediately follows that the spin moments agree with the matrix quantum mechanics answer (\ref{eq:largemoments}). Given that the large $N$ limit taken in the matrix quantum mechanics did not presuppose $N = 2^k$, this seems to indicate that classical states with energy close to the crystalline states always exist in the large $N$ limit, although we do not know how to find them.

\subsection{Existence of the topological quantum phase transition}
\label{sec:exist}

Let us explicitly make the argument that the matrix saddle undergoes a continuous large $N$ topological quantum phase transition of the sort described in section \ref{sec:transition} above.

We have seen that the matrix saddle (\ref{eq:Z3}) of the bosonic partition function correctly captures the ground state to several orders in perturbation theory about the free limit $v_4 = 0$ (section \ref{sec:smallv}) and in the classical limit $v_4 = \infty$ (section \ref{sec:largev}). Away from these limits, the mismatch described in section \ref{sec:smallv} suggests that the higher derivative terms in the expansion of the discrete derivative (\ref{eq:ddt}) may be important. However, these terms are still consistent with an emergent $O(N)$ symmetry. Such terms would lead to additional contributions to perturbation theory but do not lead to a breakdown of the ansatz (\ref{eq:relax}) for the Lagrange multipliers. The $O(N)$ symmetry implies that a collective field description will exist at all couplings. The eigenvalue distribution is connected at $v_4 = 0$ and disconnected at $v_4 = \infty$, so there must be a topological phase transition at intermediate couplings. Because the eigenvalue distribution at a given coupling is unique (this is at least true for perturbative corrections to the free and classical limits), the transition will be continuous and in the universality class of that described in section \ref{sec:transition} above.

The conclusion above pertains to the matrix saddle, independently of additional physics such as quantum glassiness and ordering that may dominate regions of the zero temperature phase diagram.

\section{Low energy excitations}
\label{sec:excitations}

We can now characterize the gapless excitations at and close to the continuous quantum critical coupling $v_4^\text{QCP}$. These will be shown to be described by waves propagating in an emergent 1+1 dimensional spacetime.

\subsection{Collective excitations of the eigenvalue distribution}

The collective field Hamiltonian (\ref{eq:col2}) can be written in the more transparent form
\be
H=\int d\lambda\rho(\lambda)\left( \frac{1}{2K}\left[\left[(P\partial\pi)(\lambda)\right]^2+\frac{\pi^{2}}{3}\frac{1}{4}\rho(\lambda)^{2}\right]+V(\lambda)\right) -\frac{N^{2}}{32K} \,.
\ee
Here the constraint (\ref{eq:twoc}) has been used as well as the expression (\ref{eq:simp}) to simplify the Hilbert transform. We have introduced the projection of the momentum
\be
(P\partial\pi)(\lambda)=\int d\lambda' \r(\l') P(\lambda,\lambda')\partial_{\lambda'}\pi(\lambda') \,,
\ee
where
\be
P(\lambda,\lambda')=\frac{\delta(\lambda-\lambda')}{\rho(\lambda')}-\frac{\lambda \lambda'}{N^{2}} \,.
\ee
The constraint (\ref{eq:twoc}) means that $P(\lambda,\lambda')$ satisfies the properties of a projection operator. 

The collective eigenvalue excitations are found by considering small perturbations around the ground state solution
\begin{eqnarray}
\rho(\lambda,t) & = & \rho_\star(\lambda)+\delta\rho(\lambda,t) \,, \\
\pi(\lambda,t) & = & 0+\delta\pi(\lambda,t) \,.
\end{eqnarray}
The linearized perturbations are controlled by the quadratic Hamiltonian
\be\label{eq:HH2}
H^{(2)}=\frac{1}{2K} \int d\lambda\rho_\star(\lambda)\left( \left[(P_\star\partial\delta\pi)(\lambda)\right]^2+\frac{\pi^{2}}{4}\delta\rho(\lambda)^{2} \right) \,. 
\ee
Here $P_\star$ denotes the projector evaluated on the background solution $\rho_\star$. 

Following \cite{Mondello:1980du, dasjev} there are two
things we can do to simplify the quadratic Hamiltonian (\ref{eq:HH2}). Firstly, 
the change of variables to $q(\l)$, such that
\be
\frac{d\lambda}{dq} = \rho_\star(\l) \,.
\ee
Secondly, the canonical transformation to new variables $\{\phi,\pi_\phi\}$:
\begin{eqnarray}
\delta\rho & = & \frac{1}{\rho_\star} \sqrt{\frac{2}{\pi}} \, \partial_{q}\phi \,,  \label{eq:dphi} \\
\partial_{\lambda}\delta\pi & = & - \frac{1}{\rho_\star} \sqrt{\frac{\pi}{2}} \, \pi_\phi \,.
\end{eqnarray}
The quadratic Hamiltonian then takes a more conventional form
\be\label{eq:H2nice}
H^{(2)}=\frac{\pi}{4 K}\int dq\left( (\bar{P}\pi_\phi)^{2}+(\partial_{q}\phi)^{2}\right) \,,
\ee
where the projector in the new variables $q$ can be written as
\be
\bar{P}(q,q')=\delta(q-q')-\frac{1}{4 N^{2}}\partial_{q}(\lambda^{2}(q))\partial_{q'}(\lambda^{2}(q')) \,.
\ee

The change of variables above has the additional benefit of turning one of the constraints into a boundary condition.
Let the variable $q$ run from $-L$ to $L$, with
\be\label{eq:LL}
2 L = \int_{-L}^L dq = \int \frac{d\l}{\r_\star(\l)} \,.
\ee
Then we can write the constraint (that the total number of eigenvalues does not change) as
\be
0 = \int \delta \rho \, d \lambda = \sqrt{\frac{2}{\pi}} \int \pa_q \phi dq = \sqrt{\frac{2}{\pi}} \left(\phi(L) - \phi(-L) \right) \,.
\ee
Choosing the constant of integration from (\ref{eq:dphi}) in the most natural way, so that $\phi(-L)=0$, one therefore has the boundary condition
\be\label{eq:bc}
\phi(-L) = \phi(L) = 0 \,.
\ee

The equations above are quite similar to the well-known results for the $c=1$ matrix model, e.g. \cite{Mondello:1980du, dasjev}, and in particular one sees an emergent 1+1 dimensional free field dynamics in the quadratic Hamiltonian (\ref{eq:H2nice}). The new ingredient is the remaining constraint, the linearization of the second constraint in (\ref{eq:twoc}), which can be written
\be\label{eq:consQ}
Q \equiv \int_{-L}^L dq \, \partial_{q}(\lambda^{2}(q)) \, \phi(q) = 0 \,.
\ee
This constraint commutes with the quadratic Hamiltonian (\ref{eq:H2nice}), due to the presence of the projection operator, as it should. This is a nonlocal constraint. To see the extent to which the emergent collective eigenvalue dynamics is local, we proceed to explicitly solve the equations of motion and characterize the linear modes.

\subsection{Solution to the perturbation equations}

The Hamiltonian equations of motion can be cast in the following form
\bea
\dot{\phi} & = & \frac{\pi}{2 K} \left(\pi_\phi -  \g \, \partial_{q}(\lambda^{2}) \right) \,, \\
\dot{\pi}_\phi & = & \frac{\pi}{2 K}\partial_{q}^{2}\phi\,, 
\eea
where $\g$ (which depends on time but not $q$) is to be fixed by imposing that the solution obey the constraint (\ref{eq:consQ}) at all times. These equations can be solved. Looking for solutions with a definite frequency of oscillation, and imposing part of the boundary conditions, $\phi(-L) = 0$, one has
\bea
\g & = & e^{- i \omega t} i \g_o \,, \label{eq:gg} \\
\phi & = & e^{- i \omega t} \left[ \phi_o \sin \frac{\omega (q+L)}{c} + \g_o \int_{-L}^q \sin \frac{\omega (q-q')}{c} \pa_{q'}(\l^2(q')) dq' \right] \,, \label{eq:pp}
\eea
where
\be\label{eq:cc}
c = \frac{\pi}{2K} \,.
\ee
The overall normalization is unimportant. The remaining boundary condition, $\phi(L)=0$, and the constraint (\ref{eq:consQ}) will fix the ratio $\g_o/\phi_o$ and will quantize the frequencies $\omega$.

The above solution simplifies when
\be\label{eq:wlarge}
\omega \gg \frac{h}{L} \,,
\ee
which we will now show is also the limit in which a local dispersion is recovered. In the solution, the frequency appears in the combination $\omega L/c$; to see the factor of $L$, rescale the coordinate $q = L \bar q$ in (\ref{eq:pp}), which eliminates all other factors of $L$ from the expression. Recall that $c$ is related to $K$ via (\ref{eq:cc}), which in turn
depends on $h$ via (\ref{eq:Kh}). Thus we can think in terms of the combination $\omega L/h$. The condition (\ref{eq:wlarge}) will then be the statement that a large number of microsopic spins participate in the collective mode. In the limit (\ref{eq:wlarge}) of large $\omega$, integration by parts, together with careful treatment of boundary terms -- in particular, using the fact that near the endpoints of the eigenvalue distribution $\rho_\star(\lambda) \sim |\l-\l_\pm|^{1/2}$ -- shows that the integral in (\ref{eq:pp})
\be
\int_{-L}^q \sin \frac{\omega (q-q')}{c} \pa_{q'}(\l^2(q')) dq' = \frac{c \, \partial_{q}(\lambda^{2}(q))}{\omega} + \ocal\left(\frac{1}{\omega^2}\right) \,.
\ee
This is the usual cancellation in a rapidly oscillating integral.
Using this result and (\ref{eq:pp}) in the constraint (\ref{eq:consQ}), and performing similar integrations by parts, one finds that at large $\omega$
\be
\frac{\gamma_o}{\phi_o} =  \ocal\left(\frac{1}{\omega}\right) \,.
\ee
The previous two equations together imply that the second term in (\ref{eq:pp}), that contains the effects of the nonlocal constraint, drops out at large $\omega$. In this limit, the solution that further satisfies the remaining boundary condition $\phi(L) = 0$ is then
\be\label{eq:wn1}
\phi(t,q) = \phi_o e^{- i \omega_n t} \sin \frac{\omega_n (q+L)}{c} \,, \qquad \omega_n = \frac{n \, c \, \pi}{2 L} \,, \qquad n \gg 1 \,,
\ee
with $n$ integer.
These are just the solutions to the free wave equation
\be\label{eq:scalar}
\ddot \phi - c^2 \pa^2_q \phi = 0 \,,
\ee
with the boundary conditions $\phi(-L) = \phi(L) = 0$. This, finally, is our sought-after emergent spacetime locality from the spin system. This is the wave equation we quoted in (\ref{eq:locality}) above.

The energy carried by the collective modes can be written in units of the microscopic energy scale $h$, using the definition of $c$ in (\ref{eq:cc}) as well as (\ref{eq:Kh}), 
\be\label{eq:wofh}
\omega_n =  \frac{4 \pi^2}{L} \, n \, h \,.
\ee
As the system is tuned to the critical point, the length $L$ diverges. From the definition of L in (\ref{eq:LL}), it is clear that $L$ will diverge logarithmically if the background eigenvalue distribution $\rho_\star(\l)$ vanishes linearly in $\l$ at some point. In the solution (\ref{eq:dist}) this is seen to occur at $\l=0$ when $a = 0$. For $a$ small this means that the integral will be dominated by the contribution close to $\l = 0$. Using the critical values of the parameters given in (\ref{eq:vcc}) and (\ref{eq:lcc}), one obtains
\be\label{eq:LLL}
L = \frac{5}{6} \log \, \left(\hat v_4^\text{QCP} - \hat v_4\right)^{-1} + \cdots \,,
\ee
as the quantum critical point is approached.

As $L \to \infty$ near the critical point, the condition (\ref{eq:wlarge}) is compatible with the excitations having small energy relative to the bare microscopic scale: $h/L \ll \omega \ll h$. This means that these excitations are consistently captured by the matrix quantum mechanics in which the higher derivative terms in (\ref{eq:ddt}) are dropped. They are part of the universal critical excitations. It is clear from these modes that the energy gap is closing logarithmically as the critical point is approached. A more precise bound can be put on the gap as follows.

A certain class of critical states can be found for all $n$, including order one values, as we now explain. The boundary conditions (\ref{eq:bc}) and constraint (\ref{eq:consQ}) can also be satisfied by setting
\be\label{eq:wn2}
\gamma_o = 0 \,, \qquad \omega = \omega'_n = \frac{(2 n-1) \, c \, \pi}{2 L} \,, \qquad n \in \Z^+ \,,
\ee
in the solution given by (\ref{eq:gg}) and (\ref{eq:pp}). The $\omega'_n$ correspond to odd values of $n$ in the $\omega_n$ of (\ref{eq:wn1}), but $n$ no longer needs to be large. The constraint (\ref{eq:consQ}) is solved because $\sin [\omega_n' (q+L)/c]$ is even in $q$ whereas $\pa_q (\l^2)$ is odd. These low-lying excited states allow us to bound the closure of the gap as $L \to \infty$ at the quantum critical point. This is done by putting $n=1$ in (\ref{eq:wn2}) and using the growth (\ref{eq:LLL}) in $L$ as $\hat v_4 \to \hat v_4^\text{QCP}$. The result is the bound
\be\label{eq:gap}
\frac{\Delta E_\text{min.}}{h} \leq \frac{24 \pi^2}{5} \frac{1}{\log (h/\Delta v)} \,.
\ee
on the lowest excitation close to the quantum phase transition, $\Delta v = \hat v_4^\text{QCP} - \hat v_4$.

The excitations (\ref{eq:wn2}) are only a subset of the very low energy modes. While beyond the objectives of this study, it should be possible to solve the constraint equations exactly, or at least numerically, and thereby obtain the full spectrum of excitations that are captured by the matrix quantum mechanics.

\subsection{Microscopic description of the excitations}

At any fixed distance from the critical point, the length $L$ is an order one number, and in particular has no $N$ scaling. Comparing with the microscopic Hamiltonian (\ref{eq:H}), we see that (\ref{eq:wofh}) is roughly the energy to flip $n$ spins. To get $\omega \gg h/L$ it is therefore sufficient to flip a large but order one number of spins. These are the collective excitations described by the emergent 1+1 dimensional free scalar field (\ref{eq:scalar}). Microscopically speaking, the excitations will be particular superpositions of flipped spins. We can get some limited intuition for what these superpositions are from weak coupling.

Evaluating the integral (\ref{eq:LL}) at small $\hat v_4$, using the background solution (\ref{eq:dist}), one finds
\be
L = \pi^2 \left( 1 - 32 \hat v_4 + 3200 \hat v_4^2 - 393216 \hat v_4^3 + \cdots  \right) \,.
\ee
The modes with frequencies (\ref{eq:wofh}) then lead to the excitation energies
\be\label{eq:pertDeltaE}
\frac{\Delta E_n}{h} = n \left( 4 + 8 \frac{v_4}{h} - 34 \left(\frac{v_4}{h}\right)^2 + 216\left(\frac{v_4}{h}\right)^3 + \cdots  \right) \,.
\ee
The first term in this expansion, $\Delta E^{(0)}_n = 4 h n$, has a satisfying interpretation. It is the energy cost of flipping $n$ off-diagonal spins in the free theory or alternatively $2 n$ diagonal spins (or some combination thereof). To see this, first note that, with no interactions ($v_4 = 0$), the ground state of the spin system has all spins pointing in the $x$ direction:
\beq\label{eq:psi0}
\ket{\psi_0} = \ket{\rightarrow}^{\otimes N(N+1)/2}.
\eeq
Here $\ket{\rightarrow}$ is the eigenvector of the $\sigma^x$ Pauli matrix such that
\bea
\sigma^x \ket{\rightarrow} =  \ket{\rightarrow} &\quad& \sigma^x \ket{\leftarrow} = - \ket{\leftarrow} \,, \nonumber \\
\sigma^z \ket{\rightarrow} =  \ket{\leftarrow}  &\quad& \sigma^z \ket{\leftarrow} =  \ket{\rightarrow}.
\eea
Excited states then correspond to flipping a certain number of spins from $\ket{\rightarrow}$ to $\ket{\leftarrow}$.
According to the Hamiltonian (\ref{eq:H}), flipping $n$ off-diagonal and $m$ diagonal spins costs energy
\be\label{eq:de1}
\Delta E^{(0)}_n = 2 h (2 n + m) \,.
\ee
The energy difference for a single flipped spin is $2h$, but because of the symmetry constraint, if we flip an off-diagonal spin $S_{AB}$, we must also flip $S_{BA}$. 

The excited states with energy $\Delta E_n^{(0)} = 4 h n$ are highly degenerate. At first order in perturbation theory in $v_4$, the degeneracy is partially lifted. Let us focus on states that do not primarily involve flipping diagonal spins, as in the mapping to the matrix saddle in (\ref{eq:relax}) we were not careful to treat the dynamics of the diagonal spins correctly. One set of states that reproduce the degeneracy-splitting energy of $\Delta E_n^{(1)} =8 v_4 n$ predicted by the collective field in (\ref{eq:pertDeltaE}) are as follows. These states are given by a linear superposition of all possible ways of flipping $k$ `rectangles' of off-diagonal spins (i.e. a total of $n = 4 k$ spins). For instance, for $k=1$ this would be
\be\label{eq:excitedS}
|\text{excited singlet}\rangle = |12,23,34,41\rangle + |12,23,35,51 \rangle + \cdots + |12,23,3N,N1 \rangle + \cdots \,. 
\ee
Here $|A,B,C,D\rangle$ denotes a state in which the four spins $A,B,C,D$ have been flipped relative to the ground state (\ref{eq:psi0}). This state is invariant under permuting rows and columns and is therefore a singlet state. That is good, as the collective field should indeed be describing singlet excitations. We can see that the splitting of the degenerate state indeed leads to $\Delta E_n^{(1)} =8 v_4 n$ for these states as follows. The quartic interaction Hamiltonian in (\ref{eq:H0}) -- let's call it $\delta H$ -- mixes two different rectangles of spins that share one edge with a factor of $8$. For example:
\be
\langle 12,23,34,41 |\delta H | 12,23,35,51\rangle = \frac{8 v_4}{N} \,.
\ee
Furthermore, a state with $k$ rectangles can mix with $4 k N = n N$ other different states with $k$ rectangles. This is because one can choose any of the $4$ edges of the $k$ rectangles to mix, and a given edge can mix with $N$ different other edges. The resulting restricted matrix elements of $\delta H$ admit an eigenstate given by the
excited singlet state (\ref{eq:excitedS}), with eigenvalue $\Delta E_n^{(1)} =8 v_4 n$.

States such as (\ref{eq:excitedS}) will become highly dressed at order one couplings. Nonetheless, they give some feel for the types of spin excitations involved in the emergent critical behavior.

\subsection{Non-singlet modes}

Non-singlet modes of the matrix quantum mechanics are described by the Hamiltonian (\ref{eq:Hoff}). Given a background eigenvalue distribution, this is a quadratic Hamiltonian for the matrix $\Omega$ in (\ref{eq:Omeg}) whose spectrum can be found exactly in principle. At generic points in parameter space there is no hierarchy between the energy of singlet and non-singet excitations \cite{Marchesini:1979yq}.

Components of the matrix $\Omega$ are associated with pairs of eigenvalues, as in for instance the Hamiltonian (\ref{eq:Hoff}). The modes become heavy if the corresponding eigenvalues are closely spaced. Upon tuning to the quantum critical point, there is a divergence in the density of states close to the point where the eigenvalue distribution splits. This leads to tightly spaced eigenvalues in this region and a divergence in the energy of the corresponding non-singlet excitations as \cite{Gross:1990md, Boulatov:1991xz,Maldacena:2005hi}
\be
\frac{\Delta E_\text{non-sing.}}{h} \sim \log (1/a) \sim  \log (h/\Delta v) \,.
\ee
While the decoupling only occurs close to the disconnection point of the eigenvalue distribution (cf. \cite{Maldacena:2005hi}), recall from the computation around (\ref{eq:LLL}) above that this is the region responsible for the growth of the length $L \to \infty$. Therefore, this is the same region where the gapless singlet excitations propagate locally.

An alternative approach to isolating the singlet dynamics would be to realize an emergent gauged matrix quantum mechanics, in which the non-singlet states are projected out. This may be possible by appropriately coupling the spin system to a dynamical magnetic field. The decoupling of the non-singlet modes at the critical point is equivalent to an emergent gauge symmetry in the low energy theory.

\section{Discussion}
\label{sec:discuss}

This paper has constructed a `regularization' of large $N$ matrix quantum mechanics by a finite dimensional spin Hamiltonian. The structure we have built has suggestive connections to other physical systems, as we now describe.

{\it Gauge theory:} It has been pointed out to us by Xiaoliang Qi and Zhao Yang that the spin Hamiltonian (\ref{eq:H0}) defines a $\Z_2$ gauge theory on a highly connected graph in which $N$ vertices are all connected to each other by $N(N+1)/2$ edges
(if the spins along the diagonal can be neglected, i.e. each vertex is not connected to itself, then this is a complete graph). The $\Z_2^N$ symmetry described in section \ref{sec:model} is then precisely the gauge symmetry that acts at the $N$ vertices, while the $N(N+1)/2$ spins live on the edges. The maximally connected nature of the graph allows for this gauge theoretic interpretation, despite the fact that there are order $N^2$ spins, but only order $N$ symmetries. From this point of view, the topological transition we have encountered is a cousin of familiar transitions in large $N$ gauge theories in which the eigenvalue distribution of Wilson loops becomes disconnected \cite{Gross:1980he, Wadia:1980cp}.

An exciting possibility is that gauge theories on complete graphs may give a general microscopic framework for realizing emergent matrix quantum mechanics, as the gauge symmetries pin the indices on the spin interactions in pairs, in a way that can naturally lead to matrix multiplication in a continuum limit. This may be helpful for finding qubit systems that allow an emergent matrix quantum mechanics with more than one matrix. This will be necessary to obtain a higher dimensional emergent spacetime with richer structure, such as black holes with Bekenstein-Hawking entropy and stringy substructure of spacetime leading to Ryu-Takayanagi entanglement. It may also be that the gauge theoretic, hyperoctahedral structure on its own is enough to connect directly with emergent spacetime dynamics.

{\it Scrambling:}
The paper \cite{parisi} considered the classical version of the system we have studied because it is a solvable model of a structural glass, that is, without quenched disorder. The matrix multiplication structure of the interactions was sufficient to obtain frustrated dynamics. Another feature of matrix interactions is that they are expected to be associated with `fast scrambling' quantum chaotic dynamics \cite{Sekino:2008he, Maldacena:2015waa}. Therefore, quantum spin systems that lead to emergent matrix quantum mechanics might be expected to give tractable models for fast scrambling. The currently best understood microscopic model for this behavior, the Sachdev-Ye-Kitaev model \cite{Sachdev:1992fk, kitaev, Polchinski:2016xgd, Maldacena:2016hyu}, involves quenched random interactions.  Another model with quenched disorder that may scramble quickly is the transverse-field Sherrington-Kirkpatrick model \cite{ray,Yao:2016ayk}. Matrix interactions may remove the need for quenched disorder in achieving fast scrambling, as they did for glassiness. Indeed, there may be a relation between glassiness and fast scrambling \cite{Yao:2016ayk}.

{\it Fermions:} Bosonic matrix quantum mechanics can also emerge as the low energy description of a many-body fermionic quantum mechanics with nonlocal interactions \cite{Anninos:2015eji}. In this case, the `qubit' nature of the underlying Hilbert space is due to the Grassmann statistics of fermions. The fact that fermions at different sites anticommute, however, means that even simple-looking Hamiltonians have a complicated representation as a matrix on the Hilbert space (cf. the Jordan-Wigner transform). Similar to the models we have obtained in this paper, the low energy boson matrix quantum mechanical wavefunctions obtained in \cite{Anninos:2015eji} are valued in a compact space. In fact, the general class of quantum mechanical models considered in \cite{Anninos:2015eji} (that paper focuses on $SU(2)$ invariant Heisenberg-like interactions) may possibly describe `fractionalized' spin liquid phases (see e.g. \cite{wen}) of transverse field spin models, via the change of variables
\be
\vec S_{AB} = {\bar \psi_{AC}}  \vec \sigma \, \psi_{CB} \,.
\ee
Here $\vec \sigma$ are the Pauli matrices and $\psi_{AC}$ is a large $N$ matrix of fermions. In these constructions there is typically also an emergent gauge symmetry acting on the fermion Hilbert space \cite{wen}.

{\it String theory:} The unconstrained single matrix quantum mechanics gives a nonperturbative description of two dimensional string theory \cite{Klebanov:1991qa, Ginsparg:1993is, Polchinski:1994mb, Gomis:2003vi}. It is natural to ask whether the constraint (\ref{eq:C1}) that arises from an underlying spin system admits a string theoretic interpretation. We can make one preliminary comment in this regard. If we define
\be
P_\pm(\l) = \pa_\l \pi(\l) \pm \frac{\pi}{2} \rho(\l) \,,
\ee
then the collective field Hamiltonian (\ref{eq:col2}), using (\ref{eq:simp}), becomes
\bea
\lefteqn{H=\int d\lambda\left( \frac{1}{6\pi K}\Big[P_{+}^{3}(\lambda)-P_{-}^{3}(\lambda)\Big]+\frac{1}{\pi}\Big[P_{+}(\lambda)-P_{-}(\lambda)\Big]V(\lambda)\right)} \nonumber \\
& &\quad -\frac{1}{8\pi^{2}KN^{2}}\left(\int d\lambda\lambda\Big[P_{+}^{2}(\lambda)-P_{-}^{2}(\lambda)\Big]\right)^{2} \,.
\eea
We see that the final term due to the constraint breaks the decoupling of the two chiral modes $P_\pm$ that is otherwise present in two dimensional string theory \cite{Polchinski:1991uq, Gross:1991qp}.

\section*{Acknowledgements}

We are especially grateful to Brian Swingle for some helpful comments that allowed this project to get going. It is also a pleasure to acknowledge useful discussions with Pallab Basu, Erez Berg, John Cardy, Sumit Das, Eduardo Fradkin, Jeff Harvey, Shamit Kachru, Andreas Karch, Steve Kivelson, Igor Klebanov, Natalie Paquette, Xiaoliang Qi, Subir Sachdev, Steve Shenker, Julian Sonner, David Tong and Zhao Yang.

\appendix

\section{Constrained quantization details}
\label{sec:dirac}

This appendix gives some details regarding the quantization of a constrained system that is performed in section \ref{sec:collectiveH} in the main text.

\subsection{Dirac quantization}

The starting point is the Lagrangian (\ref{eq:LL}), i.e.
\be
L =  \tr \left(\frac{K}{2} \dot \Phi^T \dot \Phi - V(\Phi) - \mu \left( \Phi^T \Phi - N \right)  - \nu \left(\Phi^T - \Phi \right) \right) \,.
\ee
Recall that $\mu$ is a single Lagrange multiplier, whereas $\nu_{AB}$ is a matrix's worth of multipliers. The momenta conjugate to the fields are clearly
\beq
\Pi^{AB} \equiv \frac{\partial L}{\partial \dot{\Phi}^{AB}} =  K \dot{\Phi}^{AB} , \qquad \Pi_{\mu}  \equiv \frac{\partial L}{\partial \dot{\mu}} = 0,  \qquad \Pi_{\nu_{AB}}  \equiv \frac{\partial L}{\partial \dot{\nu}_{AB}} = 0,
\eeq
and hence the na\"ive Hamiltonian is
\bea
H & = & \sum_{AB} \Pi^{AB}  \dot{\Phi}^{AB} - L \nonumber \\
& = &  \tr \left[ \frac{1}{2 K } \Pi^T \Pi + V(\Phi) + \mu  (\Phi^T \Phi -N) + \nu \left(\Phi^T - \Phi  \right) \right] \,.
\eea
However the primary constraints
\be
\chi^0_1 \equiv \Pi_{\mu} =0 \,, \qquad \chi^{AB}_1 \equiv \Pi_{\nu_{AB}} =0 \,,
\ee
imply that the most general Hamiltonian takes the form
\beq
H_T =   H + u^0_1 \chi^0_1 + \sum_{AB} u^{AB}_1 \chi^{AB}_1\,,
\eeq
where the $u_1$'s are arbitrary at this point.

For consistency we now have to require that the $\chi_1$'s are zero at all times:
\bea
0 &=& \dot{\chi}^0_1 = \{ \chi^0_1 , H_T \} = - \frac{\partial H}{\partial \mu} = -  \tr (\Phi^T \Phi -N) , \nonumber\\
0 &=& \dot{\chi}^{AB}_1 = \{ \chi^{AB}_1 , H_T \} = - \frac{\partial H}{\partial \nu_{AB}} = \Phi^{AB} - \Phi^{BA},
\eea
where we used the usual Poisson bracket:
\beq
\{ f , g \} \equiv \sum_i   \frac{\partial f}{\partial q_i}  \frac{\partial g}{\partial p_i} - \frac{\partial f}{\partial p_i}  \frac{\partial g}{\partial q_i}.
\eeq
It follows that we find the secondary constraints:
\be
\chi^0_2 \equiv - \tr (\Phi^T \Phi -N) = 0 \,, \qquad \chi^{AB}_2 \equiv \Phi^{AB} - \Phi^{BA} = 0 \,.
\ee
Since these constraints commute with the primary constraints, we can also add them to the Hamiltonian:
\beq
H_T =   H + u^0_1 \chi^0_1 + \sum_{AB} u^{AB}_1 \chi^{AB}_1 + u^0_2 \chi^0_2 + \sum_{AB} u^{AB}_2 \chi^{AB}_2,
\eeq
and now require that the $\chi_2$'s are zero at all times. This gives
\bea
0 &=& \dot{\chi}^0_2 = \{ \chi^0_2 , H_T \} = - \frac{2}{K} \sum_{AB} \Phi^{BA} \Pi^{BA} , \nonumber\\
0 &=& \dot{\chi}^{AB}_2 = \{ \chi^{AB}_2, H_T \} = \frac{1}{K} \left( \Pi^{AB} - \Pi^{BA} \right) \,,
\eea
and thus further secondary constraints
\be
\chi^0_3 \equiv - 2 \tr (\Phi^T \Pi) = 0 \,, \qquad \chi^{AB}_3 \equiv \Pi^{AB} - \Pi^{BA} = 0 \,.
\ee
These constraints cannot be added to the Hamiltonian without changing the dynamics of the $\chi_2$'s since they do not commute. Instead, imposing $0 = \dot{\chi}^0_3 = \{ \chi^0_3 , H_T \}$ and $0 = \dot{\chi}^{AB}_3 = \{ \chi^{AB}_3 , H_T \}$ fixes $u_2^0$ and $u^{AB}_2$, respectively, and does not introduce any further constraints.

At this stage the dynamics is still not fully determined because we still have the unknown parameters $u_1$. These free parameters indicate a gauge freedom and can be fixed by the convenient gauge conditions:
\bea
\mu = 0 &\Rightarrow& 0 = \dot{\mu} = \{ \mu , H_T \} = u^0_1, \nonumber\\
\nu_{AB} = 0 &\Rightarrow& 0 = \dot{\nu}_{AB} = \{ \nu_{AB} , H_T \} = u^{AB}_1. 
\eea
In the end we find that the system can be described by the Hamiltonian: 
\be
H = \tr \left(\frac{1}{2 K} \Pi^T \Pi + V(\Phi) \right) \,,
\ee
together with the constraints
\bea
\tr \left(\Phi^T \Phi - N\right) = 0 \,, &\quad& \tr (\Phi^T \Pi) = 0 \,, \\
\Phi - \Phi^T= 0 \,,&\quad& \Pi - \Pi^T = 0 \,.
\eea
These are the results quoted in the main text.

To ensure that observables remain on the constrained surface, the dynamics and commutators must be evaluated using the Dirac bracket:
\beq
\{ f , g \}_\text{Dirac} \equiv \{ f , g \} - \sum_{ij} \{ f , \chi_i \} C_{ij} \{ \chi_j , g \} ,
\eeq
where $C = M^{-1}$, the inverse of the Poisson bracket matrix of the constraints:
\be
M_{ij} = \{ \chi_i , \chi_j \} \,.
\ee
The eight sets of constraints are
\bea
& \chi^0_1 = \Pi_{\mu} , \qquad \chi^0_2 = - \tr (\Phi^T \Phi -N), \qquad \chi^0_3 = - 2 \tr (\Phi^T \Pi), \qquad \chi^0_4 = \mu \,, \\
& \chi^{AB}_1 = \Pi_{\nu_{AB}} , \qquad \chi^{AB}_2 =\Phi^{AB} - \Phi^{BA}, \qquad \chi^{AB}_3 = \Pi^{AB} - \Pi^{BA}, \qquad \chi^{AB}_4 = \nu_{AB}  \,.
\eea
It is straightforward to compute and invert $M$, with the result
\be
\{ \Phi^{AB} , \Pi^{CD} \}_\text{Dirac}  = \frac{1}{2}(\delta_{AC} \delta_{BD} + \delta_{AD} \delta_{BC}) -  \frac{1}{N^2} \Phi^{AB} \Phi^{CD}.
\ee
This is the result quoted in (\ref{eq:dirac}) in the main text.

\subsection{Rotation to eigenvalue basis}

In going from the Hamiltonian acting on wavefunctions of the matrix $\Phi$, equation (\ref{eq:H1}), to the eigenvalue Hamiltonian (\ref{eq:Hlam}), the following manipulation is important.

The symmetric matrix $\Phi$ can be diagonalized via $\Phi = O^T \Lambda O$, where $\Lambda_{ij} = \lambda_i \delta_{ij}$. It follows that
\beq
\delta (\Phi_{AB}) = \delta (O^T \Lambda O)_{AB} = \delta (O^T)_{AC} (\Lambda O)_{CB} +(O^T)_{AC} \delta(\lambda)_{C} O_{CB} + (O^T \Lambda)_{AC} \delta(O)_{CB},
\eeq
which implies that
\beq
\frac{\partial \Phi_{AB}}{\partial \lambda_i} = (O^T)_{Ai}  O_{iB}.
\eeq
Using this result we find
\beq\label{eq:phidphi}
\sum_i \lambda_i \frac{\partial}{\partial \lambda_i} = \sum_{ABi} \lambda_i \frac{\partial \Phi_{AB}}{\partial \lambda_i} \frac{\partial}{\partial \Phi_{AB}} = \sum_{AB} \Phi_{AB} \frac{\partial}{\partial \Phi_{AB}}.
\eeq

\section{Perturbative matching with the spin system}
\label{sec:matching}

In this appendix we match the matrix quantum mechanics expansions of the moments (\ref{eq:moments})
with perturbative computations in the microscopic model (\ref{eq:H0}). The reader may find it instructive to see how a combinatorial structure arises from the spin system.

For ease of reference, recall that the transverse-field Ising model (\ref{eq:H0}) to be solved is
\be\label{eq:H4}
H = - h \sum_{A,B} S^{\,x}_{AB} + \frac{v_4}{N} \sum_{A,B,C,D} S^z_{AB} S^z_{BC} S^z_{CD} S^z_{DA} \,.
\ee
All indices run from $1$ to $N$, although the identification (\ref{eq:S}) means that there are only $N(N+1)/2$ spins.
This model can be analyzed perturbatively in $v_4$ using textbook quantum mechanical perturbation theory. Small $v_4$ is equivalent to large $h$ and hence this limit will describe quantum disordered spins.

When $v_4 = 0$, the ground state has all spins pointing in the $x$ direction:
\beq\label{eq:psi00}
\ket{\psi_0} = \ket{\rightarrow}^{\otimes N(N+1)/2}.
\eeq
Here $\ket{\rightarrow}$ is the eigenvector of the $\sigma^x$ Pauli matrix such that
\bea
\sigma^x \ket{\rightarrow} =  \ket{\rightarrow} &\quad& \sigma^x \ket{\leftarrow} = - \ket{\leftarrow} \,, \nonumber \\
\sigma^z \ket{\rightarrow} =  \ket{\leftarrow}  &\quad& \sigma^z \ket{\leftarrow} =  \ket{\rightarrow}.
\eea

\subsection{Moments from the spin system: zeroth order}

From the $v_4 = 0$ ground state we can evaluate the moments at zeroth order in perturbation theory in $v_4$. In terms of the microscopic spin variables
\be
\langle \tr \, \Phi^{2n} \rangle = \bra{\psi_0} \tr \left[(S^z)^{2n}\right] \ket{\psi_0} \,,
\ee
with $\ket{\psi_0}$ given in (\ref{eq:psi00}), with all spins pointing in the $x$-direction. To leading order in large $N$, this expectation value is obtained by solving the following combinatorial problem: The operator $S^z$ flips a spin that is pointing along the $x$-direction. It follows that only the terms in $ \tr \left[(S^z)^{2n}\right]$ that flip any given spin an even number of times can contribute to the expectation value in the ground state. Of these terms, those that flip $n$ spins twice dominate at large $N$, that is terms such as:
\beq
\sum_{A B_1 B_2 \dots B_n} S^z_{AB_1} S^z_{B_1A} S^z_{AB_2} S^z_{B_2A} \dots S^z_{AB_n} S^z_{B_nA} =  \sum_{A B_1 B_2 \dots B_n} 1 = N^{n+1} \,, \label{eq:pairing}
\eeq
where we used $S^z_{AB} S^z_{BA} = (S^z_{AB})^2 = 1$. We thus have to count how many of these terms there are in
\beq
 \tr \left[(S^z)^{2n}\right] = \sum_{A_1 \dots A_n B_1 \dots B_n} S^z_{A_1 B_1} S^z_{B_1 A_2} S^z_{A_2 B_2} \dots S^z_{A_n B_n} S^z_{B_n A_1} \,.
\eeq

We consider the first operator, which acts on spin $A_1B_1$, and take any of the other operators and force it to also act on spin $A_1B_1$. It turns out that at large $N$ we only have to consider the cases $A_1 = A_i$ and $B_1 = B_{i-1}$ with $i=1, \dots, n$, where we defined $B_0 \equiv B_n$. That is, the cases where the second, fourth, sixth, ..., or 2n-th operator acts on the same spin as the first operator. The other cases give subleading contributions at large $N$. When we do this, i.e. enforce $A_1 = A_i$ and $B_1 = B_{i-1}$, something nice happens. There are two types of terms that arise, those in which the identified spins are adjacent in the trace and those in which they are separated. This leads to
\bea
\lefteqn{ \sum_{A_1 \dots A_n B_1 \dots B_n} \left(\sum_{i=1 \dots n} \delta_{A_1 A_i} \delta_{B_1 B_{i-1}}\right) S^z_{A_1 B_1} S^z_{B_1 A_2} S^z_{A_2 B_2} \dots S^z_{A_n B_n} S^z_{B_n A_1} }\nn
&& \qquad = \; 2 N \tr [(S^z)^{2n-2}] \; + \sum_{i=2 \dots n-1} \tr [(S^z)^{2i-2}] \tr [(S^z)^{2n-2i}] \,.
\eea
This result leads to a recursion relation. Let us define
\beq
\bra{\psi_0} \tr [(S^z)^{2n}] \ket{\psi_0} \equiv I_{2n} + \mathcal{O}(N^n) \,,
\eeq
then at leading order in large $N$ we have just shown that
\beq
I_{2n} = 2 N I_{2n-2} + \sum_{i=2, \dots n-1} I_{2i-2} I_{2n-2i}\,.
\eeq
The above recursion relation is solved by precisely the expression found from the matrix saddle in (\ref{eq:moments}):
\beq\label{eq:I2n}
I_{2n} = N^{n+1} \frac{2 \Gamma(2n)}{\Gamma(n)\Gamma(n+2)} \,.
\eeq
This agreement shows explicitly that the matrix quantum mechanics is solving the correct combinatorial problem.
In particular, the `matrix saddle' indeed captures the fully quantum disordered large $N$ ground state with $v_4/h = 0$.

\subsection{Moments from the spin system: first order}

We can now include the first order correction in $v_4$. We need to compute:
\beq
\bra{\psi_0} \tr [(S^z)^{2n}] \ket{\psi_1} = \sum_{k \neq \psi_0} \bra{\psi_0} \tr [(S^z)^{2n}] \ket{ k} \frac{\bra{ k} \d H \ket{\psi_0}}{E_0 - E_k} \,. \label{eq:p0p1}
\eeq
Here we have used the standard perturbation theory formula for the first order correction to the state, $\ket{\psi_1}$. Thus
$E_0$ and $E_k$ are the energies of $\ket{\psi_0}$ and $\ket{k}$ in the unperturbed Hamiltonian, $H_0$. $H_0$ and $\d H$ are the first and second terms in (\ref{eq:H4}).

The action of $\d H$ on the ground state is to flip 2 or 4 spins from the plus $x$ to the minus $x$ direction. It follows that the states $\ket{k}$ have only 2 or 4 spins flipped and thus $ \bra{\psi_0} \tr [(S^z)^{2n}] \ket{ k}$ is non-zero only when $ \tr [(S^z)^{2n}]$ flips back these same 2 or 4 spins and when it flips all other spins that it acts on an even number of times. At leading order in $N$ we only have to consider the terms in $ \tr [(S^z)^{2n}]$ that flip $n-2$ spins twice, and the remaining $4$ spins once. In particular, we find that the leading contributions come from terms that can be written as follows:
\beq
\sum_{A_1\dots A_4} S^z_{A_1A_2} T_{A_2} (k_1) S^z_{A_2A_3} T_{A_3} (k_2) S^z_{A_3A_4} T_{A_4} (k_3) S^z_{A_4A_1} T_{A_1} (k_4) \,, \label{eq:Ts} 
\eeq
where
\beq
T_{A}(k) \equiv (S^z)^{2k}_{AA} \,,
\eeq
and where we pair up the operators within each $T$ as in (\ref{eq:pairing}). That is, the operators in the $T$s are the ones that flip the $n-2$ spins twice. In terms of the discussion of the previous subsection, we see that to leading order
\be
T_A(k) = \frac{I_{2k}}{N} \,.
\ee
There are then four types of terms contributing to (\ref{eq:p0p1}) depending on whether the four $T$s in (\ref{eq:Ts}) are contiguous or not. Taking combinatorial factors into account we obtain:
\begin{align}
\tr [(S^z)^{2n}]  & =  \left( \frac{2n}{N} I_{2n-4} + \frac{3n}{N^2} \sum_k I_{2k} I_{2n-2k-4} + \frac{2n}{N^3} \sum_{k,m} I_{2k} I_{2m} I_{2n-2k-2m-4} \right. \nonumber \\
&  \qquad \left. + \, \frac{n}{2N^4} \sum_{j,k,m} I_{2j} I_{2k} I_{2m} I_{2n-2j-2k-2m-4} \right)  \tr [(S^z)^4]  + \cdots \nonumber \\
& =   \frac{1}{N^3} \frac{n(n-1)}{n+2} \, I_{2n} \, \tr [(S^z)^4]  + \cdots \,. \label{eq:s2n1}
\end{align}
The last step used the expression (\ref{eq:I2n}) for $I_{m}$.

Using the result (\ref{eq:s2n1}) in (\ref{eq:p0p1}), the first order in $v_4$ correction to the moment is
\bea
2 \bra{\psi_0} \tr (S^z)^{2n} \ket{\psi_1} &=& \frac{2}{N^3} \frac{n(n-1)}{n+2} I_{2n} \frac{N}{v_4} \sum_k \frac{|\bra{ k} \delta H \ket{\psi_0}|^2}{E_0 - E_k} + \cdots \nn
&=& \frac{2}{N^3} \frac{n(n-1)}{n+2} I_{2n} \frac{N}{v_4} \frac{-v_4^2 N^2}{2 h}  + \cdots \nn
&=& - \frac{n(n-1)}{n+2} I_{2n} \frac{v_4}{h}  .
\eea
This expression agrees precisely with the leading order matrix quantum mechanics correction in (\ref{eq:moments}), provided that we make the identification quoted in (\ref{eq:Kh}), namely
\be\label{eq:Krelate}
K = \frac{1}{16 \, h} \,.
\ee
Thus the matrix quantum mechanics ground state is solving the correct combinatorial problem even upon the (perturbative) inclusion of the quartic spin interactions. The entire $n$ dependence (all the single trace moments, effectively the entire eigenvalue distribution) has been reproduced from matching a single effective coupling in (\ref{eq:Krelate}).

We have similarly reproduced the order $v_4^2$ moments in (\ref{eq:moments}).

\end{document}